\documentclass[aps,prb,reprint,superscriptaddress,onecolumn,floatfix,notitlepage]{revtex4-1}

\usepackage{amsmath}
\usepackage{amssymb}
\usepackage{amsfonts}
\usepackage{natbib}
\usepackage{graphicx}

\newcommand{\ket}[1]{\left| #1\right\rangle}

\newcommand{\bra}[1]{\left\langle #1 \right|}
\newcommand{\an}[1]{d_{#1}^{}}
\newcommand{\dda}[1]{d^{\dagger}_{#1}}
\newcommand{\chup}{\chi_{\uparrow}}
\newcommand{\chd}{\chi_{\downarrow}}
\newcommand{\cho}{\chi_0}
\newcommand{\zb}{\bar{z}}
\newcommand{\wb}{\overline{w}}
\newcommand{\ex}[1]{\langle #1 \rangle}
\newcommand{\nor}[1]{:\mathrel{#1}:}

\newcommand{\dnb}{\partial}

\newcommand{\ntot}{n^{\mathrm{tot}}}
\newcommand{\msum}{\sideset{}{'}\sum}
\newcommand{\jb}{\bar{j}}
\newcommand{\kb}{\bar{k}}

\newcommand{\ocft}{\mathcal{O}_{\mathrm{CFT}}^{\lambda}}

\newcommand{\hb}{\bar{h}}
\newcommand{\nb}{\bar{n}}

\newcommand{\mv}[1]{\mathbf{#1}}

\DeclareMathOperator{\sgn}{sgn}
\DeclareMathOperator{\tr}{Tr}

\begin{document}
	
\title{Parent Hamiltonians for lattice Halperin states\\from free-boson conformal field theories}

\author{Anna Hackenbroich}
\affiliation{Max-Planck-Institut f\"ur Quantenoptik, Hans-Kopfermann-Stra{\ss}e 1, D-85748 Garching, Germany}
\author{Hong-Hao Tu}
\email{h.tu@lmu.de}
\affiliation{Max-Planck-Institut f\"ur Quantenoptik, Hans-Kopfermann-Stra{\ss}e 1, D-85748 Garching, Germany}
\affiliation{Physics Department, Arnold Sommerfeld Center for Theoretical Physics and Center for NanoScience,
Ludwig-Maximilians-Universit\"at M\"unchen, 80333 M\"unchen, Germany}

\begin{abstract}
	We introduce a family of many-body quantum states that describe interacting spin one-half hard-core particles with bosonic or fermionic statistics on arbitrary one- and two-dimensional lattices. The wave functions at lattice filling fraction $\nu=2/(2m+1)$ are derived from deformations of the Wess-Zumino-Witten model $\mathfrak{su}(3)_1$ and are related to the $(m+1,m+1,m)$ Halperin fractional quantum Hall states. We derive long-range SU(2) invariant parent Hamiltonians for these states which in two dimensions are chiral $t$-$J$-$V$ models with additional three-body interaction terms. In one dimension we obtain a generalisation to open chains of a periodic inverse-square $t$-$J$-$V$ model proposed in [Z.\ N.\ C.\ Ha and F.\ D.\ M.\ Haldane, Phys.\ Rev.\ B \textbf{46}, 9359 (1992)]. We observe that the gapless low-energy spectrum of this model and its open-boundary generalisation can be described by rapidity sets with the same generalised Pauli exclusion principle. A two-component compactified free boson conformal field theory is identified that has the same central charge and scaling dimensions as the periodic bosonic inverse-square $t$-$J$-$V$ model.
\end{abstract}

\maketitle	

\section{Introduction}
Two-dimensional conformal field theory (CFT) is a valuable tool for the analysis of a large class of strongly correlated quantum systems in one and two spatial dimensions. CFT may be used to describe the gapless edge modes of two-dimensional systems with chiral topological order such as quantum Hall samples~\cite{Klitzing1980,Tsui1982,Laughlin1983} and chiral spin liquids~\cite{Kalmeyer1987,Wen1989}. Moreover, the low-energy effective theory for quantum critical systems of spins or fermions on one-dimensional lattices links the critical exponents of correlation functions to the scaling dimensions of a CFT. As first noted by Moore and Read~\cite{Moore1991} CFT may furthermore be used to construct the many-body wave functions for the ground state and elementary quasi-hole excitations in the two-dimensional bulk of fractional quantum Hall (FQH) systems. In a similar spirit it was suggested to use the correlation functions of conformal fields to define many-body wave functions for one- and two-dimensional lattice states~\cite{Cirac2010,Nielsen2012,Tu2015obc,Montes2016}. These states are referred to as infinite matrix product states (MPS) due to a formal similarity to usual MPS constructed from finite-dimensional matrices. For infinite MPS wave functions constructed from a rational CFT, long-ranged lattice parent Hamiltonians can be derived that possess the infinite MPS as their exact ground state~\cite{Cirac2010,Nielsen_su2_k} (there exist alternative ways of deriving similar parent Hamiltonians, see, e.g., Refs.~\onlinecite{schroeter2007,greiter2014}). On one-dimensional chains with periodic or open boundary conditions one thus obtains quantum critical chains~\cite{Cirac2010,Nielsen_su2_k,Tu2013,Tu2014,Tu2014_sun,Bondesan2014,Tu2015obc} such as the Haldane-Shastry (HS) model~\cite{Haldane1988,Shastry1988}, whereas on generic two-dimensional lattices the construction yields chiral topological states~\cite{Nielsen2012,Glasser2015,Glasser2016}. In most but not all cases, the CFT characterising these one- or two-dimensional phases is closely related to the theory which defines the many-body wave functions of the infinite MPS.\par
The clarification and elucidation of the phase diagram of cuprate high-temperature superconductors is one of the biggest and most long-standing open problems in theoretical condensed matter physics~\cite{Lee2006}. These systems are usually studied using the $t$-$J$ model~\cite{zhang1988} which is the strong-coupling limit of the single-band Hubbard model and describes itinerant spin one-half fermions without double occupancy of any lattice site that interact through spin exchange. On one-dimensional chains, the Hubbard and $t$-$J$ models are gapless quantum critical Tomonaga-Luttinger liquids~\cite{Haldane1981} whose low-energy effective CFT can be constructed from free-boson theories. In 1992 Ha and Haldane introduced certain long-range $t$-$J$-$V$ models with modified density-density interaction for bosonic or fermionic particles defined on periodic one-dimensional chains without double occupancy and with interaction strengths decaying as the inverse square chord distance~\cite{ha1992}. They constructed a set of low-energy eigenstates with lattice wave functions very similar to the spin-singlet $(m+1,m+1,m)$ Halperin FQH state~\cite{Halperin1983} which is the most natural generalisation to spin-unpolarised systems of the Laughlin state~\cite{Laughlin1983}\label{key} at filling $1/(m+1)$. Since the lattice analogue of the simplest bosonic Laughlin state at filling 1/2 is just the ground state of the SU(2) HS model, the infinite MPS construction based on the CFT $\mathfrak{su}(2)_1$ provides a direct relation between the spin-polarised FQH state and the one-dimensional lattice model for spin one-half particles without holes. One may ask whether a similar connection exists between spin-singlet Halperin FQH samples and the one-dimensional quantum critical $t$-$J$-$V$ models from Ref.~\onlinecite{ha1992} for hole-doped spin one-half systems.\par
In this paper we identify a two-dimensional chiral CFT such that the infinite MPS derived from this theory essentially provides this link between Ha and Haldane's inverse-square $t$-$J$-$V$ models and the Halperin spin-singlet FQH wave function. From the correlator of fields from this CFT we construct on arbitrary one- and two-dimensional lattices a spin-singlet state at lattice filling fraction $2/(2m+1)$ with a Jastrow wave function identical to the $(m+1,m+1,m)$ Halperin state. We derive long-range SU(2) invariant parent Hamiltonians for this infinite MPS on generic lattices which describe interacting spin one-half hard-core bosons (fermions) for odd (even) values of $m$. In two dimensions, we thereby obtain a chiral $t$-$J$-$V$ model with additional three-body interaction terms. In one-dimension the result provides a generalisation of Ha and Haldane's model to chains with open boundary conditions, whereas the parent Hamiltonian on periodic chains differs from the inverse-square $t$-$J$-$V$ model only by an additional term that explicitly breaks time reversal symmetry. Using Monte-Carlo calculations we analyse the entanglement entropy and correlation functions in the Halperin infinite MPS on periodic and open chains and find that the states are quantum critical and described by a low-energy CFT with central charge $c=2$. Moreover, we observe that the distinct energy levels in the gapless finite-size spectrum of Ha and Haldane's inverse-square $t$-$J$-$V$ model and its open-boundary generalisation can be described by rapidity sets similarly to the HS model~\cite{Haldane1988} but with a generalised Pauli exclusion principle~\cite{Haldane1991}. From the finite-size scaling of the resulting analytic expressions for the energy and momentum of the low-energy states we extract the conformal dimensions of the primary fields in the low-energy CFT of the periodic model. We identify the action of a toroidally compactified two-component free boson CFT~\cite{Sule2015} whose partition function on the torus precisely agrees with the observed spectrum of scaling dimensions in the bosonic periodic model at odd values of $m$.\par
The paper is organised as follows. In Sec.~\ref{sec:setup}, we define an infinite MPS from a CFT of two free bosons and use the algebraic structure of this theory to construct lattice operators that annihilate the state on arbitrary one- and two-dimensional lattices. In Sec.~\ref{sec:iMPSwf}, we analyse the nature of the infinite MPS, first by deriving the form of its wave function on different lattices and then by numerically studying the entanglement entropy and two-point correlation functions on periodic and open chains. In Sec.~\ref{sec:Hamil}, explicit expressions for the infinite MPS parent Hamiltonians in one and two dimensions are provided. Furthermore, we suggest a description for the finite-size spectrum of the periodic and open inverse-square $t$-$J$-$V$ Hamiltonians in terms of rapidity sets, derive the lowest scaling dimensions of the periodic models and identify a two-component free boson CFT matching the periodic bosonic models. Finally, we conclude the paper by mentioning some possible directions for future research in Sec.~\ref{sec:conclusion}.

\section{Constructing models for spin one-half hardcore particles from free-boson CFTs\label{sec:setup}}

\subsection{Infinite MPS for spin one-half hard-core bosons or fermions}
We consider interacting hard-core particles of species $\uparrow$ or $\downarrow$ and with bosonic or fermionic statistics moving on a lattice $\Gamma=\{z_i\in\mathbb{C}|i=1,\dotsc,N\}$ embedded into the complex plane. Each lattice site can be either empty $\ket{0}$ or occupied by a particle $\ket{\sigma}$ of species $\sigma=\,\uparrow,\downarrow$, whereas double-occupancy configurations are excluded from the Hilbert space. We propose an ansatz state
\begin{equation}
\label{iMPS}
\ket{\psi}=\sum_{\mu_1,\dotsc,\mu_N=0,\uparrow,\downarrow}\psi(\mu_1,\dotsc,\mu_N;z_1,\dotsc,z_N)\ket{\mu_1,\dotsc,\mu_N}
\end{equation}defined by a lattice wave function
\begin{equation}
\label{CFTwavefunction}
\psi(\mu_1,\dotsc,\mu_N;z_1,\dotsc,z_N)=\left\langle A^{\mu_1}(z_1)\dotsm A^{\mu_N}(z_N)\right\rangle
\end{equation}which is the expectation value of a product of conformal fields $A^{\mu}$ evaluated at the positions of the lattice sites $z_1,\dotsc,z_N$. Just as for translation-invariant MPS, the operator $A^{\mu_i}$ inserted at the $i^{\text{th}}$ position in the correlation function giving the coefficient of the state $\ket{\mu_1,\dotsc,\mu_N}$ depends only on the configuration $\ket{\mu_i}$ of the $i^{\text{th}}$ lattice site. Since the Hilbert space of a two-dimensional CFT is infinite-dimensional, \eqref{iMPS} is sometimes referred to as an infinite MPS. It is fully determined by a choice of three conformal fields $A^{0,\uparrow,\downarrow}$, one for each local basis state. In previous work it was established that infinite MPS characterising systems of spin one-half particles without holes are based on the CFT $\mathfrak{su}(2)_1$~\cite{Cirac2010}. Moreover, systems of spin-less particles at filling fractions $\nu=1/q$ less than unity can be described by infinite MPS derived from vertex operators of a chiral free boson compactified at radius $\sqrt{q}$~\cite{Tu2014}. In order to describe spin one-half particles at filling fractions less than unity we combine these two observations and consider the family of chiral vertex operators
\begin{equation}\label{defA}
A^{\mu}(z)=\chi_{\mu}e^{is_{\mu}\phi_1/\sqrt{2}}\,e^{i((2m+1)n_{\mu}-2)\phi_2/\sqrt{2(2m+1)}}
\end{equation}parametrised by an integer number $m\in\mathbb{N}$. Here, the parameters $n_0=0$, $n_{\uparrow,\downarrow}=1$ and $s_0=0$, $s_{\uparrow,\downarrow}=\pm1$ characterise the occupation number and spin of a single site in the three different basis states. The operators~\eqref{defA} are elements of a CFT with central charge $c=2$ of two chiral real massless free bosons $\phi_1, \phi_2$ compactified at the radii $R_1=\sqrt{2}$ and $R_2=\sqrt{2(2m+1)}$ that describe the spin and charge degree-of-freedom of the hard-core particles, respectively. This CFT contains six current operators that define a closed chiral algebra with respect to which the vertex operators $A^{\mu}$ form the three components of a primary field (we use the term 'current' for the elements of a chiral algebra irrespective of whether their conformal dimension is $h=1$~\cite{Schellekens_CFT}). The only singular term in the operator product expansion (OPE) of any one of these currents $\mathcal{O}_{\mathrm{CFT}}$ with the fields $A^{\mu}$ is given by
\begin{equation}\label{cftrep}
\mathcal{O}_{\mathrm{CFT}}(z)A^{\mu}(w)=-\sum_{\nu}\frac{(\mathcal{O})_{\mu\nu}}{z-w}A^{\nu}(w),
\end{equation}where $(\mathcal{O})_{\mu\nu}$ is the representation matrix of a single-site linear operator of the hardcore-particle lattice system (see Appendix~\ref{app:cft} for an explanation of the normalisation convention). In this sense, each CFT current is linked to a local lattice operator and the algebraic structure of the free-boson CFT reflects the structure of the local Hilbert space and operator algebra of the lattice system. The resulting map between the conformal fields and the lattice operators or states is summarised in Tab.~\ref{tab:cftrep}. Denoting by $\dda{\sigma}$ the operator that creates a hard-core boson or fermion of species $\sigma$, the lattice SU(2) spin generators $S^a=\frac{1}{2}\sum_{\alpha,\beta=\uparrow,\downarrow}\dda{\alpha}\sigma_{\alpha\beta}^a\an{\beta}$ for $a=1,2,3$ and $S^{\pm}=S^{1}\pm iS^{2}$ correspond to the currents $J^3_1(z)=-(i/\sqrt{2})\dnb \phi_1(z)$ and $J^{\pm}_1(z)=\pm\chup\chd e^{\mp i\sqrt{2}\phi_1(z)}$ that form an $\mathfrak{su}(2)_1$ Ka\v{c}-Moody algebra in the sector of the first boson $\phi_1$. On the other hand, the U(1) current $J_2(z)=i\sqrt{\frac{2m+1}{2}}\dnb\phi_2(z)$ of the second boson $\phi_2$ is associated with the particle number operator $n=\dda{\uparrow}\an{\uparrow}+\dda{\downarrow}\an{\downarrow}$. Finally, the particle annihilation operators $\an{\sigma}$ are represented by two currents $B^{\sigma}(z)=\chi_0\chi_{\sigma} e^{s_{\sigma}i\frac{1}{\sqrt{2}}\phi_1(z)}e^{i\sqrt{\frac{2m+1}{2}}\phi_2(z)}$ that mix the sectors of both free bosons. The vertex operators are multiplied by representations of anti-commuting Klein factors $\{\chi_{\mu},\chi_{\nu}\}=2\delta_{\mu\nu}$ in terms of Majorana fermions $\chi_{\mu}^{\dagger}=\chi_{\mu}$ which ensure the correct statistical phase for the conformal currents and primaries: Whereas $J^a_1,J_2$ with conformal dimension $h=1$ commute with the primaries $A^{\mu}$, the currents $B^{\sigma}$ with conformal dimension $h=(m+1)/2$ (anti-)commute with the two components $A^{\uparrow,\downarrow}$ of non-zero particle number for odd (even) $m$. Hence, the algebraic structure of the CFT corresponds to a bosonic (fermionic) lattice system for odd (even) $m$. A special case that has been covered in previous work~\cite{Tu2014_sun,Bondesan2014} arises for $m=1$, when all currents have conformal dimension $h=1$ and correspond to six of the eight generators of the Wess-Zumino-Witten (WZW) model $\mathfrak{su}(3)_1$ where $A^{\mu}$ form the three components of the WZW primary field associated with the fundamental representation of $\mathfrak{su}(3)$.

\begin{table}[tb]
	\centering
	\begin{tabular}{c c c }
		local lattice operator or state & conformal field & $h$\\
		\hline
		$S^3$ & $J^3_1=-\frac{i}{\sqrt{2}}\dnb \phi_1$ & $1$\\
		$S^{\pm}$ & $J^{\pm}_1=\pm\chup\chd e^{\mp i\sqrt{2}\phi_1}$ & $1$\\
		$1-\frac{2m+1}{2}n$ & $J_2=i\sqrt{\frac{2m+1}{2}}\dnb\phi_2$ & $1$\\
		$\an{\sigma}$ & $B^{\sigma}=\cho\chi_{\sigma} e^{is_{\sigma}\frac{1}{\sqrt{2}}\phi_1}e^{i\sqrt{\frac{2m+1}{2}}\phi_2}$ & $\frac{m+1}{2}$\\
		$\ket{0}$ & $A^0=\cho e^{-i\frac{2}{\sqrt{2(2m+1)}}\phi_2}$ & $\frac{1}{2m+1}$\\
		$\ket{\sigma}$ & $A^{\sigma}=\chi_{\sigma}\, e^{is_{\sigma}\frac{1}{\sqrt{2}}\phi_1}e^{i\frac{2m-1}{\sqrt{2(2m+1)}}\phi_2}$ & $\frac{m-1}{2}+\frac{1}{2m+1}$
	\end{tabular}
	\caption{Representation of the local Hilbert space and local operator algebra of a system of spin one-half hardcore particles in terms of primary fields and currents of a $c=2$ free boson CFT. For $m$ odd (even) the CFT maps to the bosonic (fermionic) lattice system.}
	\label{tab:cftrep}
\end{table}

\subsection{Null fields}
The Hilbert space of the CFT generated by the currents $\mathcal{O}_{\mathrm{CFT}}$ when acting on the primary $A^{\mu}$ contains null states that have vanishing overlap with all other states. Since the wave function~\eqref{CFTwavefunction} is given as the expectation value of product of primary fields the null states and their associated null fields may be used to derive operators which annihilate the infinite MPS and which can be combined to form a parent Hamiltonian~\cite{Nielsen_su2_k}. The simplest null fields for the infinite MPS~\eqref{iMPS} are obtained by an expansion of the product $\mathcal{O}_{\mathrm{CFT}}(z)A^{\mu}(w)$ to order $(z-w)^0$. In the $\mathfrak{su}(2)_1$ sector of the first boson $\phi_1$ there exist four null fields at the first Virasoro level~\cite{Nielsen_su2_k}. For the derivation of SU(2) invariant parent Hamiltonians it is convenient~\cite{Nielsen_su2_k} to consider three linear combinations which are labelled by a vector index $a=1,2,3$ and are given as
\begin{equation}\label{nullfieldspin}
	\lambda^{a}(w)=\frac{1}{2\pi i}\oint_{\mathcal{C}_w}\frac{dz}{z-w}[J^a_1(z)\sum_{\alpha=\uparrow,\downarrow}A^{\alpha}(w)+i\sum_{b,c=1,2,3}\epsilon_{abc}J^c_1(z) \sum_{\alpha,\beta=\uparrow,\downarrow}\frac{1}{2}(\sigma^b)_{\beta\alpha}A^{\alpha}(w)].
\end{equation}Here, $\mathcal{C}_w$ is an integration contour that circles the point $w$ once in the positive sense. In addition, we find four null fields that involve degrees-of-freedom from both $\phi_1$ and $\phi_2$
\begin{subequations}\label{nullfieldcharge}
	\begin{gather}
	\omega^{\sigma}(w)=\frac{1}{2\pi i}\oint_{\mathcal{C}_w}\frac{dz}{z-w}\left[ B^{\sigma}(z)A^0(w)-(s_{\sigma}J_1^3(z)-J_2(z))A^{\sigma}(w)\right],\\
	\eta^{\sigma}(w)=\frac{1}{2\pi i}\oint_{\mathcal{C}_w}\frac{dz}{z-w}B^{\sigma}(z)A^{\sigma}(w).
	\end{gather}There are two further operators
	\begin{gather}
	\zeta^{\sigma}(w)=\frac{1}{2\pi i}\oint_{\mathcal{C}_w}\frac{dz}{z-w}B^{-\sigma}(z)A^{\sigma}(w)
	\end{gather}
\end{subequations}which are null fields of all CFTs with $m\geq 2$.

\subsection{Operators annihilating the lattice Halperin state\label{sec:OpAnn}}

In this subsection we follow Ref.~\onlinecite{Nielsen_su2_k} and compute operators that annihilate the infinite MPS~\eqref{iMPS} on arbitrary one- or two-dimensional lattices. The simplest descendant fields generated by the action of the currents $\mathcal{O}_{\mathrm{CFT}}\in\{J_1^a,J_2,B^{\sigma}\}$ on the primary $A^{\mu}$ are linear combinations
\begin{equation}\label{nullfield}
\phi(w)=\sum_{\lambda}\phi_{\lambda}(w)=\sum_{\lambda}\frac{1}{2\pi i}\oint_{\mathcal{C}_w}dz\;f_{\lambda}(z;w)\;\mathcal{O}_{\mathrm{CFT}}^{\lambda}(z)\,A^{\alpha_{\lambda}}(w)
\end{equation}with a meromorphic scalar function $f_{\lambda}(z;w)=\sum_{n\leq -1}c_{\lambda,n}(z-w)^n$ and $\alpha_{\lambda}\in\{0,\uparrow,\downarrow\}$ for all $\lambda$. We denote by $\mathcal{O}^{\lambda}_i$ the linear operator on the lattice site $i$ that is associated with the current $\ocft$ according to Tab.~\ref{tab:cftrep}. We assume that it is possible to find a local basis state $\ket{\beta}\in\{\ket{\uparrow},\ket{\downarrow},\ket{0}\}$ such that for every term in the linear combination~\eqref{nullfield} the local lattice operator $\mathcal{Q}^{\lambda}=\ket{\beta}\bra{\alpha_{\lambda}}$ is bosonic (fermionic) for bosonic (fermionic) $\mathcal{O}^{\lambda}$. Below we show that whenever $\phi$ is a null field, the state~\eqref{iMPS} is annihilated for any value $j\in\{1,\dotsc,N\}$ by the operator
\begin{equation}\label{annop}
\Phi_j\equiv\sum_{\lambda}\Phi^{\lambda}_j\equiv\sum_{\lambda}\sum_{k(\neq j)} f_{\lambda}(z_k;z_j)\mathcal{Q}^{\lambda}_j\mathcal{O}^{\lambda}_k=\sum_{\lambda}\sum_{k(\neq j)} f_{\lambda}(z_k;z_j)(\ket{\beta}\bra{\alpha_{\lambda}})_j\;\mathcal{O}^{\lambda}_k,
\end{equation}where we denote by $j_1,\dotsc,j_n\,(\neq i_1,\dotsc,i_k)$ the set of indices $j_1,\dotsc,j_n\in\{1,\dotsc,N\}\setminus\{i_1,\dotsc,i_k\}$. The null fields~\eqref{nullfieldspin} and~\eqref{nullfieldcharge} are of the form~\eqref{nullfield} with $f_{\lambda}(z;w)\propto 1/(z-w)$. Moreover, for each of these fields it is possible to find a basis vector $\ket{\beta}\in\{\ket{\uparrow},\ket{\downarrow},\ket{0}\}$ such that the grading of the operators $\mathcal{O}^{\lambda}$ and $\mathcal{Q}^{\lambda}$ is identical for all terms $\lambda$ appearing in its definition. Therefore we can use the result~\eqref{annop} to construct their associated lattice operators which annihilate the infinite MPS. These operators will be used in Sec.~\ref{sec:Hamil} to build parent Hamiltonians for the state~\eqref{iMPS}.\par
As a first step in the calculation relating the null field~\eqref{nullfield} and the lattice operator~\eqref{annop} we insert the component $\phi_{\lambda}(z_j)$ in place of the operator $A^{\mu_j}(z_j)$ into the wave function~\eqref{CFTwavefunction} at position $z_j$. The integral over $z$ that appears in the null field then acts on the integrand $f(z;z_j)\ex{\ocft(z)A^{\mu_1}(z_1)\dotsm A^{\mu_N}(z_N)}\vert_{\mu_j=\alpha_{\lambda}}$ which is holomorphic everywhere except in the points $z=z_k$ for $k=1,\dotsc,N$, where it has poles. Using the theorem of residues, the integral along the curve $\mathcal{C}_{z_j}$ circling $z_j$ may be transformed into the sum of a positive integral over a circle with infinite radius and integrals with negative orientation circling the points $z=z_k$ for $k\neq j$, such that
\begin{gather}
\nonumber  \ex{A^{\mu_1}(z_1)\dotsm A^{\mu_{j-1}}(z_{j-1})\phi_{\lambda}(z_j) A^{\mu_{j+1}}(z_{j+1})\dotsm A^{\mu_N}(z_N)}=\\
\nonumber (-1)^{2h_{\lambda}\sum_{l=1}^{j-1}n(\mu_l)}\times\lim_{R\rightarrow\infty}\;\frac{1}{2\pi i}\oint_{|z|=R}dz\; f_{\lambda}(z;z_j)\;\ex{\mathcal{O}_{\mathrm{CFT}}^{\lambda}(z)A^{\mu_1}(z_1)\dotsc A^{\mu_N}(z_N)}\bigr\vert_{\mu_j=\alpha_{\lambda}}\\
\label{nfstep1}
-(-1)^{2h_{\lambda}\sum_{l=1}^{j-1}n(\mu_l)}\times\sum_{k(\neq j)} \frac{1}{2\pi i}\oint_{\mathcal{C}_{z_k}}dz\; f_{\lambda}(z;z_j)\ex{\mathcal{O}_{\mathrm{CFT}}^{\lambda}(z)A^{\mu_1}(z_1)\dotsm A^{\mu_N}(z_N)}\bigr\vert_{\mu_j=\alpha_{\lambda}}.
\end{gather}Here, $h_{\lambda}$ is the conformal dimension of the field $\mathcal{O}_{\mathrm{CFT}}$ such that the phase factors in~\eqref{nfstep1} account for the minus signs that appear when a fermionic current $B^{\sigma}$ is commuted past a primary $A^{\sigma}$ associated with a non-zero number of particles. Explicit evaluation of the integrand in the second line of~\eqref{nfstep1} following~\eqref{corrvertexop} and~\eqref{corru1wardid} shows that it decays faster that $|z|^{-2}$; hence the integral vanishes in the limit $R\rightarrow\infty$. The integrals over $\mathcal{C}_{z_k}$ can be simplified using the OPE~\eqref{cftrep} of the product $\mathcal{O}^{\lambda}_{\mathrm{CFT}}(z)A^{\mu_k}(z_k)$. Since the function $f_{\lambda}(z;z_j)$ is holomorphic in the vicinity of a point $z_k$ with $k\neq j$ only the first singular term $\propto (z-z_k)^{-1}$ of the OPE contributes to the integral and we find
\begin{gather}
\nonumber  \ex{A^{\mu_1}(z_1)\dotsm A^{\mu_{j-1}}(z_{j-1})\phi_{\lambda}(z_j) A^{\mu_{j+1}}(z_{j+1})\dotsm A^{\mu_N}(z_N)}=\\
\label{nfstep2}
(-1)^{2h_{\lambda}\sum_{l=1}^{j-1}n(\mu_l)}\sum_{k(\neq j)} (-1)^{2h_{\lambda}\sum_{l=1}^{k-1}n(\mu_l)\vert_{\mu_j=\alpha_{\lambda}}} f_{\lambda}(z_k,z_j) \sum_{\mu}(\mathcal{O}^{\lambda})_{\mu_k\mu}\ex{A^{\mu_1}(z_1)\dotsm A^{\mu}(z_k)\dotsm A^{\mu_N}(z_N)}\bigr\vert_{\mu_j=\alpha_{\lambda}}.
\end{gather}Here, the second phase factor appears because the current $\ocft$ needs to be commuted past the primaries $A^{\mu_1}(z_1)\dotsm A^{\mu_{k-1}}(z_{k-1})$ before the OPE can be applied. The constraint $\mu_j=\alpha_{\lambda}$ in~\eqref{nfstep2} can be incorporated by acting on the infinite MPS with the operator $\mathcal{Q}^{\lambda}_j$ which annihilates all configurations for which the $j^{\text{th}}$ site is not in the state $\alpha_{\lambda}$. After summing the contributions of all configurations $\ket{\mu_1,\dotsc,\mu_N}$, \eqref{nfstep2} thus implies that
\begin{align}\label{nfstep3}
\Phi_j^{\lambda}\ket{\psi}=\sum_{k(\neq j)}f(z_k;z_j)\mathcal{Q}_j^{\lambda}\mathcal{O}_k^{\lambda}\ket{\psi}=\sum_{\{\mu_i\}}\delta_{\mu_j\beta}\ex{A^{\mu_1}(z_1)\dotsm\phi_{\lambda}(z_j)\dotsm A^{\mu_N}}\ket{\mu_1,\dotsc,\mu_N}.
\end{align}The phase factors in~\eqref{nfstep2} are compensated by minus signs that appear when the operators $\mathcal{Q}_j^{\lambda},\mathcal{O}_k^{\lambda}$ are commuted past the particle creation operators contained in the many-body basis states $\ket{\mu_1,\dotsc,\mu_N}$. This cancellation is possible because bosonic (fermionic) local lattice operators are represented by bosonic (fermionic) CFT currents and since the grading of $\mathcal{Q}^{\lambda}$ is identical to that of $\mathcal{O}^{\lambda}$. Since $\beta$ is identical for all terms in~\eqref{annop}, after a summation over $\lambda$ all terms on the left side of~\eqref{nfstep3} contain the expectation value $\ex{A^{\mu_1}(z_1)\dotsm\phi(z_j)\dotsm A^{\mu_N}}$. This correlation function is identically zero whenever $\phi$ is a null field of the CFT such that the infinite MPS is annihilated by the operator~\eqref{annop}.

\subsection{Global symmetries of the infinite MPS\label{sec:iMPSSymmProp}}
The Ward identities for the currents $\mathcal{O}_{\mathrm{CFT}}$ generating the CFT used to define the conformal wave function~\eqref{CFTwavefunction} determine the behaviour of the infinite MPS under certain global symmetries such as the spin or particle number. If $\mathcal{O}_{\mathrm{CFT}}$ denotes one of the currents $J^a_1$, $J_2$ or $B^{\sigma}$, the integral along a curve at infinity over its expectation value with any product of primaries $A^{\mu}$ vanishes,
\begin{equation}\label{Wardid}
0=\lim_{R\rightarrow\infty}\;\frac{1}{2\pi i}\oint_{|z|=R}dz\;\ex{\mathcal{O}_{\mathrm{CFT}}(z)A^{\mu_1}(z_1)\dotsc A^{\mu_N}(z_N)}.
\end{equation}For the currents $J^{\pm}_1$ and $B^{\sigma}$ this follows by scaling arguments from the explicit expression~\eqref{corrvertexop} for the expectation value of a product of vertex operators, whereas in case of $J_1^3,J_2$ it is a direct consequence of the U(1) Ward identity~\eqref{corru1wardid} for the two free bosons $\phi_1,\phi_2$~\cite{DiFrancesco}. In analogy to the calculation presented above in Sec.~\ref{sec:OpAnn}, we can use the theorem of residues to deform the integration contour at infinity to a sum over curves with opposite orientation circling the points $z_j$ for $j=1,\dotsc,N$. Each of these integrals can be evaluated using the OPE~\eqref{cftrep}. The introduction of an operator $\mathcal{Q}^{\lambda}$ is not necessary here since the correlator in~\eqref{Wardid} is not subject to any constraint of the form $\mu_j=\alpha_{\lambda}$. After summing over all contributions $\ket{\mu_1,\dotsc,\mu_N}$ the identity~\eqref{Wardid} thus implies that the infinite MPS is annihilated by $\mathcal{O}_{\mathrm{tot}}=\sum_{j=1}^N\mathcal{O}_j$, where $\mathcal{O}$ is the lattice operator associated to the current $\mathcal{O}_{\mathrm{CFT}}$ according to Tab.~\ref{tab:cftrep}. When applied to the three $\mathfrak{su}(2)_1$ currents $J^a_1$ of the boson $\phi_1$ this result implies that the infinite MPS~\eqref{iMPS} transforms in the singlet representation of the total SU(2) spin operators, $S^{a}_{\mathrm{tot}}\ket{\psi}=0$. In particular this shows that all configurations in the infinite MPS have the same number of particles of either spin species. For the U(1) current $J_2$ of the second free boson $\phi_2$, we obtain that the infinite MPS contains a fixed number of particles $M$ that is related to the number of lattice sites by the filling fraction~\footnote{This can also be seen explicitly from the invariance of the infinite MPS wave function~\eqref{CFTwavefunction} under the global U(1) symmetry of the boson $\phi_2$, which implies that the correlator vanishes unless the sum of all phases $\sum_{i=1}^N[(2m+1)n_{\mu_i}-2]=0$, or in other words $(2m+1)M-2N=0$ for any configuration with non-zero weight.}
\begin{eqnarray}
\nu=\frac{M}{N}=\frac{2}{2m+1}.
\end{eqnarray}Finally, for the currents $B^{\sigma}$ we find that the infinite MPS~\eqref{iMPS} is annihilated by the sum $\sum_{j=1}^N\an{j\sigma}$ of all annihilation operators of either spin species.

\section{Halperin states on one- and two-dimensional lattices}\label{sec:iMPSwf}

\subsection{Many-body wave function\label{sec:iMPSwavefunction}}

Since it is the expectation value of a product of free boson vertex operators the lattice wave function~\eqref{CFTwavefunction} can be evaluated explicitly using~\eqref{corrvertexop} to give
\begin{equation}\label{corr1}
\chi_{\mu_1}\dotsm\chi_{\mu_N}\prod_{i=1}^N(-1)^{(n_i-1)(i-1)}
\prod_{i=1}^N(f_N(z_i))^{n_i}\prod_{1\leq i<j\leq N}(z_i-z_j)^{\frac{1}{2}s_is_j+\frac{2m+1}{2}n_in_j}.
\end{equation}The function $f_N(z_i)=\prod_{j=1,j\neq i}^N(z_i-z_j)^{-1}$ has a closed analytic form for certain one-dimensional lattices~\cite{Nielsen_su2_k}. In order to eliminate from the wave function~\eqref{corr1} any explicit reference to the coordinates of the unoccupied lattice sites, we represent the configuration $\mu_1,\dotsc,\mu_N$ by the list of sites $x_1,\dotsc,x_{M/2}\in \{1,\dotsc,N\}$ occupied by particles of species $\uparrow$ and the list of sites $y_1,\dotsc,y_{M/2}\in \{1,\dotsc,N\}$ occupied by particles of species $\downarrow$. The uniqueness of this notation is ensured by demanding that $x_1<x_2<\dotso <x_{M/2}$ and $y_1<y_2<\dotso <y_{M/2}$. The product of Klein factors can be reordered to give $\prod_i(-1)^{(i-1)(n_i-1)}\chi_0^{N_0}\chi_{\uparrow}^{N_{\uparrow}}\chi_{\downarrow}^{N_{\downarrow}}\sgn(x_1,\dotsc,x_{N_{\uparrow}},y_1,\dotsc,y_{N_{\downarrow}})$ up to a configuration-independent factor. Here, $\sgn$ denotes the sign function that gives a minus sign whenever there is a particle of species $\downarrow$ on a lattice site with index $y_j$ that is lower $y_j < x_i$ than the index $x_i$ of a site occupied by a particle of species $\uparrow$. In particular, the Klein operator $\chi_0^{N_0}\chi_{\uparrow}^{N_{\uparrow}}\chi_{\downarrow}^{N_{\downarrow}}$ is the same for every basis state $\ket{\mu_1,\dotsc,\mu_N}$ with a non-vanishing contribution to the infinite MPS and will henceforth be dropped~\footnote{{More formally, we choose some vector $|v\rangle$ of the Klein Hilbert space with $\langle v|\chi_0^{N_0}\chi_{\uparrow}^{N_{\uparrow}}\chi_{\downarrow}^{N_{\downarrow}}|v\rangle\neq 0$ and take the expectation value in the infinite MPS wave function~\eqref{CFTwavefunction} w.r.t. the tensor product $|v\rangle\otimes|0\rangle$, where $|0\rangle$ denotes the CFT vacuum state.}}. This justifies our approach of representing the Klein factors as Majorana fermions~\cite{SenechalBos,DelftBos}. The exponent of the last factor in~\eqref{corr1} is equal to $m+1$ if both site $i$ and site $j$ are occupied by particles of the same species, $m$ if they are occupied by particles of different species and vanishes if either site is empty. Therefore the hole coordinates naturally cancel from this term and the wave function~\eqref{CFTwavefunction} becomes
\begin{equation}\label{Halperinwf}
\big(\sgn(x_1,\dotsc,x_{M/2},y_1,\dotsc,y_{M/2})\big)^{m+1}\;\prod_{i=1}^{M/2}\big[f_N(z_{x_i})f_N(z_{y_i})\big]
\prod_{1\leq k<l\leq M/2}\big[(z_{x_k}-z_{x_l})^{m+1}(z_{y_k}-z_{y_l})^{m+1}\big]\prod_{k,l=1}^{M/2}(z_{x_k}-z_{y_l})^m.
\end{equation}The last three factors are precisely the Jastrow part of the wave function for an $(m+1,m+1,m)$ double layer Halperin FQH state~\cite{Halperin1983} where the positions $\{z_i\}$ of the particles are restricted to the lattice $\Gamma$. Since the exchange of the positions of two identical particles introduces a sign $(-1)^{m+1}$, the wave function describes bosonic particles for odd $m$ and fermionic particles for even $m$. Note that the sign $\big(\sgn(x_1,\dotsc,x_{N_{\uparrow}},y_1,\dotsc,y_{N_{\downarrow}})\big)^{m+1}$ in~\eqref{Halperinwf} can be absorbed by switching to a Hilbert space basis for which the creation operators are ordered according to the species of particle they create.

\subsubsection{Halperin states on a two-dimensional lattice}

The similarities between the Halperin wave function and the lattice wave function~\eqref{Halperinwf} extend beyond the Jastrow part if the lattice $\Gamma$ is genuinely two-dimensional and can be embedded into a disc $\{z\in\mathbb{C}|\,\vert z\vert <R\}$ with radius $R$ in such a way that the area of the region closest to any lattice site $z_j$ is the same for all lattice sites. In this case, $f_N(z_j)$ converges to $e^{-|z_j|^2/4}\times e^{-i\Im[\sum_{k(\neq j)}\log(z_j-z_k)]}$ in the thermodynamic limit $N\rightarrow\infty$~\cite{Tu2014}. This convergence is fast enough that the approximate expression can be used even for moderately large lattices~\cite{Tu2014}. Hence, the wave function of the infinite MPS in the thermodynamic limit is given by
\begin{equation}\label{latticeHalperin}
e^{-\frac{1}{4}\sum_{i=1}^{M/2}[|z_{x_i}|^2+|z_{y_i}|^2]}
\prod_{1\leq k<l\leq M/2}(z_{x_k}-z_{x_l})^{m+1}\prod_{1\leq k<l\leq M/2}(z_{y_k}-z_{y_l})^{m+1}\prod_{k,l=1}^{M/2}(z_{x_k}-z_{y_l})^m
\end{equation}up to phase factors that may be absorbed into the definition of the many-body basis. This is the expression for a double-layer $(m+1,m+1,m)$ Halperin FQH state where the positions of the particles are restricted to lie on the lattice $\Gamma$. By analogy with the continuum Halperin states we expect that the infinite MPS~\eqref{iMPS} on two-dimensional lattices is a chiral spin liquid with abelian anyonic excitations.

\subsubsection{Wave function on the uniform periodic chain}

If the system is defined on a uniform chain $\Gamma=\{z_j=e^{2\pi ij/N}|j=1,\dotsc,N\}$ with periodic boundary conditions one may show that $f_N(z_j)=z_j/N$ and the infinite MPS wave function becomes
\begin{equation}\label{wavefunction1dpbc}
\prod_{i=1}^{M/2}z_{x_i}z_{y_i}\prod_{1\leq k<l\leq M/2}(z_{x_k}-z_{x_l})^{m+1}\prod_{1\leq k<l\leq M/2}(z_{y_k}-z_{y_l})^{m+1}\prod_{k,l=1}^{M/2}(z_{x_k}-z_{y_l})^m
\end{equation}up to phase factors that may be absorbed into the definition of the many-body basis states. This wave function has eigenvalue $e^{2\pi i(-m+1)/(2m+1)}$ under a lattice translation by one site along the circle. For $m\geq 2$, it therefore has a non-vanishing momentum and is not invariant under time reversal. In 1992, Ha and Haldane studied lattice wave functions for spin one-half bosons or fermions on a uniform periodic chain without double occupancy which differ from~\eqref{wavefunction1dpbc} only in the first factor $\prod_{i=1}^{M/2}z_{x_i}^{J_{\uparrow}}z_{y_i}^{J_{\downarrow}}$~\cite{ha1992}. They showed that these states are gapless low-energy eigenstates of the inverse-square $t$-$J$-$V$ Hamiltonian
\begin{equation}\label{hamha}
H_m=\frac{\pi^2}{N^2}\mathcal{P}\Big[\sum_{i\neq j}\sin^{-2}\big(\frac{\pi}{N}(i-j)\big)\big[\sum_{\sigma}c_{i\sigma}^{\dagger}c^{}_{j\sigma}+\frac{2m^2+m}{4}n_in_j+m\vec{S}_i\cdot\vec{S}_j\big]\Big]\mathcal{P},
\end{equation}in the sector of vanishing $z$-component of the total spin and filling fraction $\nu=2/(2m+1)$ provided that the positive integers $J_{\uparrow}, J_{\downarrow}$ satisfy $0\leq J_{\sigma}\leq m+1$ and $-1\leq J_{\uparrow}-J_{\downarrow}\leq 1$~\cite{ha1992}. For odd $m$ the Hamiltonian~\eqref{hamha} has a non-degenerate ground state given by the state with parameters $J_{\uparrow}=J_{\downarrow}=(m+1)/2$~\cite{ha1992}. On the other hand, for $m$ even the states with $J_{\uparrow}=J_{\downarrow}=m/2$ and $J_{\uparrow}=J_{\downarrow}=m/2+1$ are degenerate ground states~\cite{ha1992}. Hence for $m=1$ the infinite MPS is the ground state of the Hamiltonian~\eqref{hamha} which at this parameter value is identical to the SU(3) Haldane-Shastry (HS) model in agreement with previous work~\cite{Tu2014_sun,Bondesan2014} linking infinite MPS based on the WZW model $\mathfrak{su}(n)_1$ to the SU($n$) HS model. For $m\geq 2$, the infinite MPS on a uniform periodic chain is one of the low-energy states of the inverse-square $t$-$J$-$V$ model~\eqref{hamha} and differs from the ground state by local unitary transformations. Hence, diagonal observables such as the entanglement entropy and $S^3$-spin or density correlation functions are identical in the two states.

\subsubsection{Wave function on the uniform open chain}

One-dimensional quantum critical spin chains with open boundary conditions can be described by infinite MPS defined on lattices $\Gamma=\{u_j=\cos\theta_j|\theta_j\in[0,\pi]\,\forall j=1,\dotsc,N\}$~\cite{Tu2015obc}. The parent Hamiltonian of the $\mathfrak{su}(n)_1$ infinite MPS is a uniform open SU($n$) HS model when defined on three types of uniform open chains obtained as projections onto the real axis of uniform periodic chains~\cite{Tu2015obc} and moreover remains integrable on a two-parameter family of open chains~\cite{Basu-Mallick2016}. We study the Halperin infinite MPS on the uniform open chain of type I that is given by the set of angles $\theta_j=\pi(j-1/2)/N$ and for which one finds $f_N(u_j)=(-1)^{j+1}(2^N/2N)\sin\theta_j$~\cite{Tu2015obc}. Since this expression is real the infinite MPS~\eqref{iMPS} on a uniform type-I open chain is invariant under time reversal. As expected, it reduces to the ground state of the open uniform SU(3) HS model for $m=1$.

\subsection{Properties of the states on one-dimensional lattices\label{sec:Prop}}

On one-dimensional lattices the Halperin infinite MPS~\eqref{iMPS} is expected to describe a quantum critical Luttinger liquid based on its relation to the gapless SU(3) HS model for $m=1$ and the properties of infinite MPS derived from other CFTs~\cite{Nielsen_su2_k,Tu2013,Tu2014_sun,Bondesan2014}. In this subsection, we present numerical results for the entanglement entropy and two-point correlation functions that confirm the criticality of the states and in the case of periodic boundary conditions allow us to determine the central charge and certain scaling dimensions characterising the low-energy Luttinger CFT.

\subsubsection{Renyi entanglement entropy}

The leading term in the $n^{\text{th}}$ Renyi entanglement entropy (REE) $S^{(n)}(\ell)=(1-n)^{-1}\log\tr\rho^n_{\ell}$ of a quantum critical chain depends on the central charge $c$ of the low-energy effective CFT in a universal fashion~\cite{Holzhey1994,Vidal2003,Calabrese2004,Calabrese2009}
\begin{equation}\label{entcft}
S_{\mathrm{log}}^{(n)}(\ell)=\frac{c}{6\eta}\,\Big(1+\frac{1}{n}\Big)\log\left[\frac{\eta N}{\pi}\sin\frac{\pi \ell}{N}\right]+c'_2.
\end{equation}Here, $\rho_{\ell}$ is the reduced density matrix of the first $\ell$ lattice sites, $N$ is the total number of lattice sites in the chain, $c'_2$ is a non-universal constant and $\eta=1 (2)$ for periodic (open) boundary conditions at the edges of the system. There are many systems both with periodic and open boundary conditions in which the leading CFT prediction~\eqref{entcft} is obscured by subleading terms with large and possibly oscillating amplitudes. Some of these corrections decay with a critical exponent related to the scaling dimension of a relevant or irrelevant operator in the low-energy effective CFT~\cite{Laflorencie2006,Calabrese2010prl,Cardy2010,Xavier2012}. The sub-leading contribution to the $n$\textsuperscript{th} REE associated with a primary field of scaling dimension $\Delta$ is expected to be~\cite{Calabrese2010prl,Xavier2012}
\begin{equation}\label{entosc}
S^{(n)}(\ell)=F\Big(\frac{\ell}{N}\Big)\;\cos(\kappa \ell+\omega)\left[ \frac{2\eta N}{\pi}\sin \frac{\pi \ell}{N}\right]^{-2\Delta/(\eta \,n)},
\end{equation}where $F$ is an a-priori unknown function believed to be universal and $\kappa$ and $\omega$ are model-dependent parameters determining the frequency and phase of oscillations, respectively. In single-component Luttinger liquids, the leading contribution~\eqref{entosc} decays with a critical exponent $\Delta=K$ equal to the Luttinger parameter $K$~\cite{Calabrese2010prl}. To the best of our knowledge it is not fully clear which primary fields contribute to the REE in this way for more complicated critical systems. It was observed by a comparison of different models that the dominant correction of the form~\eqref{entosc} appears to be associated with the energy operator~\cite{Cardy2010,Xavier2012}, whereas studies in SU($n$) critical chains found evidence of contributions associated with all primary fields of the low-energy CFT~\cite{Demidio2015}.

\begin{figure}[t]
	\centering
	\includegraphics[width=\textwidth]{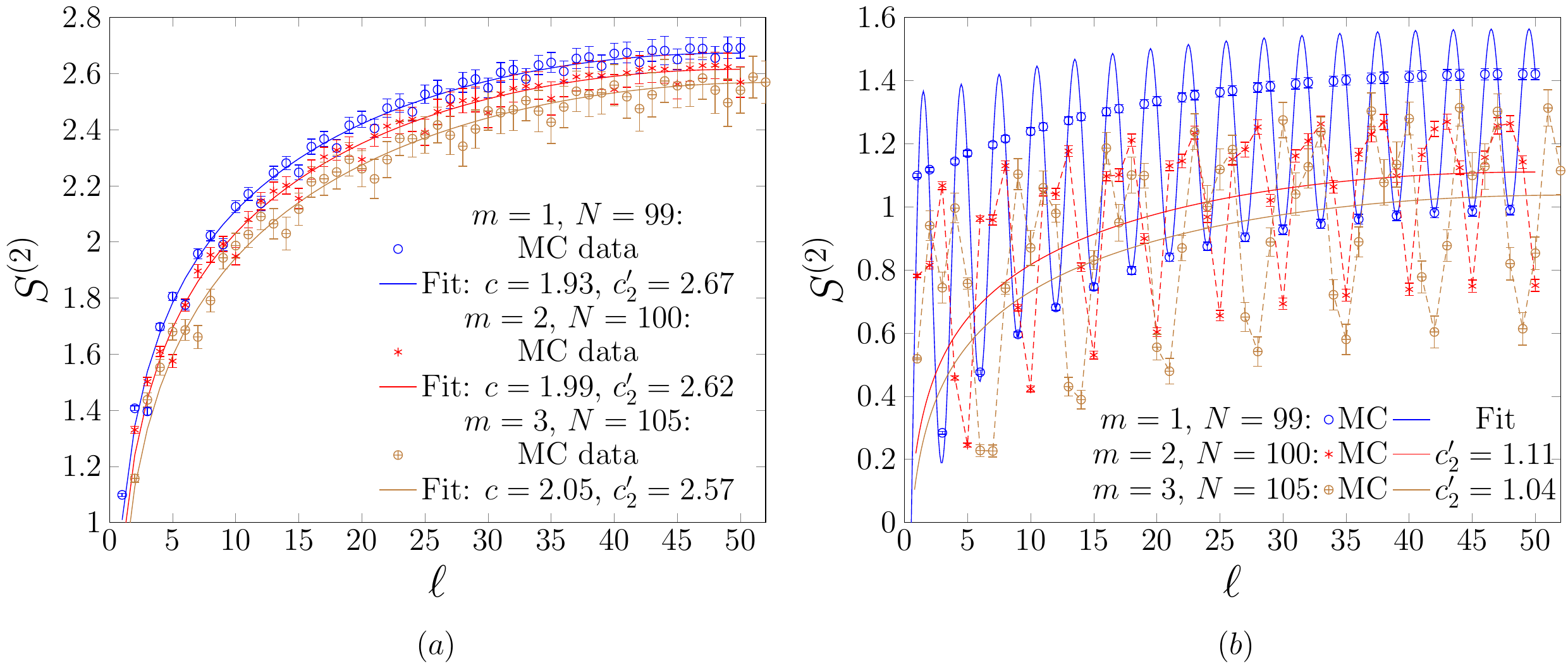}
	\caption[Renyi entropy]{Second REE $S^{(2)}(\ell)$ of the first $\ell$ sites in the Halperin infinite MPS for $m=1,2,3$ and system sizes $N=99,100,105$ obtained from Monte Carlo computations. $(a)$: The data for uniform periodic chains is fit against the leading order term~\eqref{entcft}. $(b)$: Results on the type-I uniform open chain. The data for $m=1$ is fit against the sum of \eqref{entcft} and \eqref{entosc} with $\kappa=2k_F$, $\omega=0$ and optimal parameter values $c=1.68$, $c'_2=1.28$, $4NF(\ell/N)/\pi=-0.29$, $\Delta=0.61$. For $m=2,3$ the oscillations have a complicated phase structure that cannot be reproduced by a single sub-leading term~\eqref{entosc}, but the mean of the data is well described by~\eqref{entcft} with $c=2$. The dashed lines are guides to the eye.}
	\label{fig:ent}
\end{figure}

We computed the second REE in the Halperin infinite MPS~\eqref{iMPS} using the Monte Carlo Metropolis algorithm and the replica trick~\cite{Hastings2010,Cirac2010,Tu2014}. The results for the three lowest values $m=1,2,3$ on the uniform periodic chain and the type-I uniform open chain are displayed in Fig.~\ref{fig:ent}. For both periodic and open boundary conditions sub-leading oscillatory terms at $m$ different frequencies $2qk_F$ are visible in the Fourier transform of the REE, where $q=1,\dotsc,m$ and $k_F=\pi/(2m+1)$ is the Fermi momentum. This indicates that the low-energy CFT describing the infinite MPS contains primary fields with at least $m$ different scaling dimensions. For $m=1$ in particular there are signatures of only one oscillation frequency $2k_F$; this is in agreement with analytical results, since the CFT $\mathfrak{su}(3)_1$ describing the SU(3) HS model~\cite{Kawakami1992} has two non-trivial primary fields with identical scaling dimensions $\Delta_{\mathfrak{su}(3)_1}=2/3$. For periodic boundary conditions the amplitude of these oscillatory terms is small, such that the numerical data is very well described by the leading CFT prediction~\eqref{entcft} with the central charge $c$ and the constant $c'_2$ as free parameters. We obtain values of $c$ that are very close to $c=2$ which is the expected result for the SU(3) HS at $m=1$ and furthermore agrees with the central charge of the free-boson CFT used to construct the Halperin infinite MPS~\eqref{iMPS}. For open boundary conditions the amplitude of the sub-leading oscillatory terms in the REE is much greater, in accordance with observations by other authors in different critical quantum chains~\cite{Laflorencie2006,Demidio2015}. The numerical data corresponding to the open boundary SU(3) HS model $m=1$ is well described by the sum of the leading CFT prediction~\eqref{entcft} and a sub-leading oscillatory contribution~\eqref{entosc} with constant $F$. This yields the value $c=1.68$ for the central charge, which is rather far from the expected result $c=2$ for $\mathfrak{su}(3)_1$. The discrepancy may be due to additional non-oscillatory finite-size corrections that are not contained in our fit function. However, the best-fit value for the scaling dimension $\Delta_{fit}=0.61$ is very close to the analytical result $\Delta_{\mathfrak{su}(3)_1}=2/3$. For $m=2$ and $m=3$ the REE displays oscillations without any clear phase structure which cannot be described by a single term~\eqref{entosc}. Nonetheless, there is a qualitative agreement between the mean of the numerical data and the leading CFT prediction~\eqref{entcft} with central charge $c=2$.

\subsubsection{Spin and density correlation functions}

\begin{figure}[t]
	\centering
	\includegraphics[width=\textwidth]{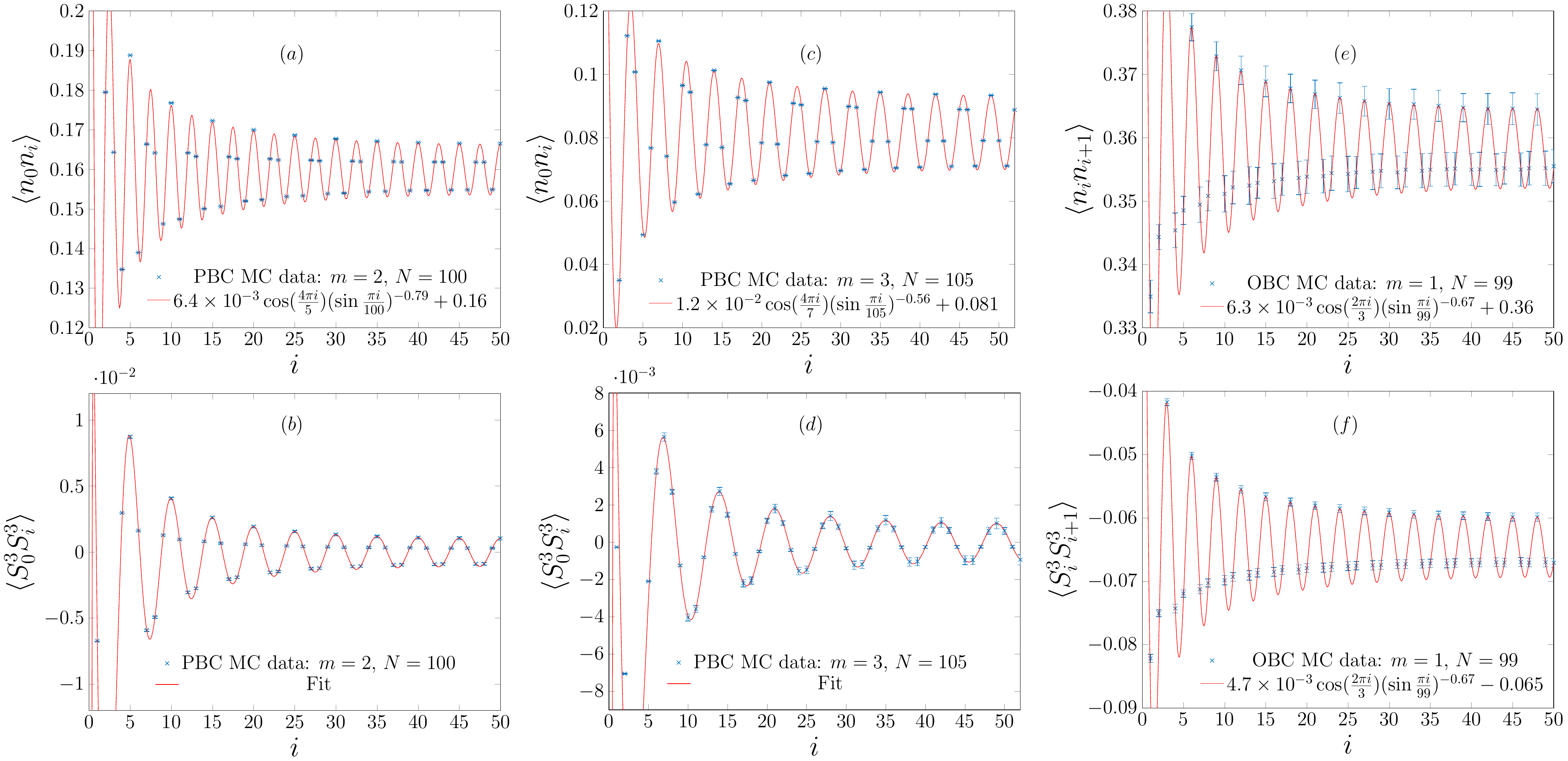}
	\caption{Numerical data from Monte Carlo computations for the spin and density correlation functions in the Halperin infinite MPS with fits as described in the main text. $(a)$, $(b)$: Density and spin correlation function for $m=2$ on a periodic uniform chain with $N=100$ sites. The fit parameters for the spin correlation function are $A_0=0$, $A_1=0.0011$, $A_2=-2.7\times 10^{-4}$, $\Delta=0.598$. $(c)$, $(d)$: Density and spin correlation function for $m=3$ on a periodic uniform chain with $N=105$ sites. The fit parameters for the spin correlation function are $A_0=0$, $A_1=0.0010$, $A_2=-2.4\times 10^{-4}$, $\Delta=0.569$. $(e)$, $(f)$: Nearest-neighbour density and spin correlation function for $m=1$ on a type-I open uniform chain with $N=99$ sites.}
	\label{fig:corrfun}
\end{figure}

Correlation functions in critical systems decay algebraically with critical exponents that are related to the scaling dimensions of primary fields in the low-energy effective CFT. At long distances and to leading order in the inverse system size, the two-point spin $\ex{S^3_0S^3_i}$ and density $\ex{n_0n_i}$ correlation functions in a periodic critical chains as well as the nearest-neighbour correlation functions $\ex{S^3_iS^3_{i+1}}$ and $\ex{n_in_{i+1}}$ in a open critical chains are expected to be of the form~\cite{white2002}
\begin{equation}\label{corransatz}
A_0+A_1\times\cos\big(2qk_Fi\big)\Big[\sin\frac{\pi i}{N}\Big]^{-2\Delta/\eta}.
\end{equation}Here, $\eta=1(2)$ for periodic (open) boundary conditions, $A_0$ and $A_1$ are non-universal constants, $q\in\mathbb{N}$ is an integer and $\Delta$ is the scaling dimension of a primary field in the low-energy effective CFT. For the periodic spin correlation function $\ex{S^3_0S^3_i}$ we extend~\eqref{corransatz} by an additional non-oscillatory term $A_2[\sin\frac{\pi i}{N}]^{-2}$ that is expected to appear in any SU(2) symmetric model since the bosonised expression for the spin operator contains the $\mathfrak{su}(2)_1$ currents with scaling dimension $\Delta=1$~\cite{Gogolin2004}. As evident from Fig.~\ref{fig:corrfun} the spin and density correlation functions in the Halperin infinite MPS are well described by the scaling form~\eqref{corransatz}. Due to the extended SU(3) symmetry both the spin and density correlators in the periodic SU(3) HS model at $m=1$ oscillate at frequency $2k_F$ and decay with the same critical exponent $1.33\approx2\Delta_{\mathfrak{su}(3)_1}=4/3$, in complete agreement with analytical results~\cite{Kawakami1992}. For $m\geq 2$, the dominant terms in the spin and density correlation function for periodic boundary conditions oscillate at different frequencies $2k_F$ and $4k_F$, respectively. The best-fit value for the critical exponent of the density correlator is very close to the value $4/(2m+1)$, indicating that the density operator is associated to a primary field of conformal dimension $h=1/(2m+1)$. Similarly, the observed critical exponent of the leading oscillatory term in the spin correlation functions is very close to $2(m+1)/(2m+1)$ such that we expect the bosonised expression for the SU(2) spin to contain a primary field with conformal dimension $h=(m+1)/(2(2m+1))$. The nearest-neighbour spin $\ex{S^3_iS^3_{i+1}}$ and density $\ex{n_in_{i+1}}$ correlation functions in the open SU(3) HS model on a uniform type-I chain obey the scaling form~\eqref{corransatz} with critical exponent $0.66\approx\Delta_{\mathfrak{su}(3)_1}=2/3$. For $m\geq 2$, the correlation functions in the infinite MPS on open uniform chains display oscillations without any clear phase structure that prevent us from extracting any critical exponents.

\section{Models for interacting spin one-half hardcore particles from free-boson CFTs\label{sec:Hamil}}
In this section we derive self-adjoint, particle-number conserving and SU(2) invariant parent Hamiltonians for the Halperin infinite MPS~\eqref{iMPS}. On generic two-dimensional lattices, the parent Hamiltonian contains long-range two- and three-body interaction terms. For one-dimensional chains we obtain a two-body Hamiltonian that generalises the inverse-square $t$-$J$-$V$ models~\eqref{hamha} studied by Ha and Haldane. Our results demonstrate which interactions stabilise the many-body state~\eqref{iMPS} on different one- and two-dimensional lattices. Furthermore, in one dimension the determination of the nature of the elementary excitations above the ground state completes the identification of the phase described by the infinite MPS.

\subsection{Parent Hamiltonians for the infinite MPS\label{sec:ParHamExplicit}}
The computation of parent Hamiltonians is based on the existence of lattice operators annihilating the infinite MPS such as the operators $\Phi_j$ derived in Sec.~\ref{sec:OpAnn} above. Indeed, any convex combination of positive operators $\Phi^{\dagger}_j\Phi_j$ defines a parent Hamiltonian since the infinite MPS is an eigenstate of the lowest eigenvalue $E_0=0$. Meaningful parent Hamiltonians that describe all degrees-of-freedom in the system and possess the correct symmetry properties are obtained by an appropriate choice of $\Phi_j$. The operators annihilating the infinite MPS that are derived according to the prescription~\eqref{annop} from the null fields~\eqref{nullfieldspin} and~\eqref{nullfieldcharge} are of the form $\Phi_j=\sum_{\lambda}\sum_{k(\neq j)}(1/(z_k-z_j))\mathcal{Q}^{\lambda}_j\mathcal{O}^{\lambda}_k$ where $\mathcal{O}^{\lambda},\mathcal{Q}^{\lambda}\in\{d_{\sigma},S^a,n\}$ are local lattice operators. Due to the existence of various discrete Fourier sums for the quantity $w_{jk}\equiv(z_j+z_k)/(z_j-z_k)$ we construct the parent Hamiltonian on generic two-dimensional lattices and periodic chains from the operators $\Phi'_j\equiv\sum_{\lambda}\sum_{k(\neq j)}w_{jk}\mathcal{Q}^{\lambda}_j\mathcal{O}^{\lambda}_k-\sum_{\lambda}\mathcal{Q}^{\lambda}_j\mathcal{O}^{\lambda}_j$. These annihilate the infinite MPS since $2z_j/(z_k-z_j)=-1-w_{jk}$ and $\sum_{j=1}^N\mathcal{O}_j\ket{\psi}=0$ as discussed in Sec.~\ref{sec:iMPSSymmProp}. The operators obtained in this way from the null fields~\eqref{nullfieldspin} and~\eqref{nullfieldcharge} are listed in the second column of Tab.~\ref{tab:oppbc}.

\begin{table}[t]
	\centering
	\begin{tabular}{|l |l|l| }
		
		\hline
		Null field& Two-dimensional lattice, periodic chain& Open chain\\
		\hline
		$\lambda^a$ &
		$\begin{aligned}\Lambda'^{a}_j=\sum_{k(\neq j)}w_{jk}[n_jS^a_k+i\epsilon_{abc}S^b_jS^c_k]\end{aligned}$ & $\begin{aligned}\Lambda''^{a}_j=\sum_{k(\neq j)}(w_{jk}+w_{j\kb})[n_jS^a_k+i\epsilon_{abc}S^b_jS^c_k]\end{aligned}$\\	
		$\omega^{\sigma}$ & $\begin{aligned}  \Omega'& ^{\sigma}_j=\,(m-1)n_{j\sigma}+\sum_{k(\neq j)}w_{jk}[\dda{j\sigma}\an{k\sigma}\\& -n_{j\sigma}\big((m+\frac{1}{2}) \,n_{k}+s_{\sigma}S^3_k-1\big)]\end{aligned}$ & $\begin{aligned}  \Omega''& ^{\sigma}_j=\sum_{k(\neq j)}(w_{jk}+w_{j\kb})[\dda{j\sigma}\an{k\sigma}\\& -n_{j\sigma}\big((m+\frac{1}{2}) \,n_{k}+s_{\sigma}S^3_k-1\big)]\end{aligned}$\\
		$\eta^{\sigma}$ & $\begin{aligned}C'^{\sigma}_j=\sum_{k(\neq j)}w_{jk}\; \an{j\sigma}\,\an{k\sigma}\end{aligned}$ & $\begin{aligned}C''^{\sigma}_j=\sum_{k(\neq j)}(w_{jk}+w_{j\kb})\; \an{j\sigma}\,\an{k\sigma}\end{aligned}$\\
		$\zeta^{\sigma}$ $(m\geq2)$& $\begin{aligned}D'^{\sigma}_j=\sum_{k(\neq j)}w_{jk}\; \an{j\sigma}\,\an{k,-\sigma}\end{aligned}$ & $\begin{aligned}D_j''^{\sigma}=\sum_{k(\neq j)}(w_{jk}+w_{j\kb})\; \an{j\sigma}\,\an{k,-\sigma}\end{aligned}$\\
		\hline

	\end{tabular}
	\caption{Operators annihilating the infinite MPS that are used in the construction of parent Hamiltonians on different lattices. For a generic lattice including the periodic chain with lattice sites $z_j$ we use the abbreviation $w_{jk}\equiv(z_j+z_k)/(z_j-z_k)$. For an open chain $\Gamma=\{u_j=\cos\theta_j|\theta_j\in[0,\pi]\,\forall j=1,\dotsc,N\}$ we define $z_j=e^{i\theta_j}$, $w_{jk}=(z_j+z_k)/(z_j-z_k)$ and $w_{j \kb}=(z_j+\zb_k)/(z_j-\zb_k)$. The operators $D'^{\sigma}_j$ and $D''^{\sigma}_j$ annihilate the infinite MPS only for $m\geq 2$.}
	\label{tab:oppbc}
\end{table}

\subsubsection{Parent Hamiltonian on two-dimensional lattice}
For $m=1$, the infinite MPS possesses an extended SU(3) symmetry and a parent Hamiltonian on generic lattices that captures all degrees-of-freedom has been found in previous work~\cite{Tu2014_sun,Bondesan2014}. We focus on the case $m\geq 2$ for which an SU(2) invariant and particle-number conserving parent Hamiltonian that describes itinerant interacting hard-core particles is given by
\begin{align}
\notag H&=\sum_{a=1,2,3}U^{\dagger}_a\Big[\sum_{j=1}^N\sum_{\sigma=\uparrow,\downarrow}\big[\big(\Omega'^{\sigma}_j\big)^{\dagger}\Omega'^{\sigma}_j+\big(C'^{\sigma}_j\big)^{\dagger}C'^{\sigma}_j+\big(D'^{\sigma}_j\big)^{\dagger}D'^{\sigma}_j\big]\Big]U_a\\
\notag &=3(m-1)^2\ntot+\sum_{j}\mu_jn_j+\sum_{k\neq j}\big[t_{jk}\sum_{\sigma=\uparrow,\downarrow}\dda{j\sigma}\an{k\sigma}+V_{jk}n_jn_k+J_{jk}\vec{S}_j\cdot\vec{S}_k\big]\\
\label{phgen}
&+\msum_{j,k,l}\big[g^{(1)}_{jkl}n_jn_kn_l+g^{(2)}_{jkl}n_l\vec{S}_j\cdot\vec{S}_k+g^{(3)}_{jkl}n_l\sum_{\sigma=\uparrow,\downarrow}\dda{k\sigma}\an{j\sigma}+g^{(4)}_{jkl}\vec{S}_l\cdot\sum_{\alpha,\beta=\uparrow,\downarrow}\dda{k\alpha}(\vec{\sigma})_{\alpha\beta}\an{j\beta} \big].
\end{align}Here, the symbol $\msum_{j,k,l}$ denotes a sum over pairwise different indices $i,j,k\in\{1,\dotsc,N\}$ and $U_a=\exp[i\frac{\pi}{2}S^a_{\mathrm{tot}}]$ for $a=1,2,3$ refers to the global spin rotations by $\pi/2$ around the $x$-, $y$- and $z$-axes. Although the positive operators $\sum_{\sigma=\uparrow,\downarrow}(\Omega'^{\sigma}_j\big)^{\dagger}\Omega'^{\sigma}_j$, $\sum_{\sigma=\uparrow,\downarrow}\big(C'^{\sigma}_j\big)^{\dagger}C'^{\sigma}_j$ and $\sum_{\sigma=\uparrow,\downarrow}\big(D'^{\sigma}_j\big)^{\dagger}D'^{\sigma}_j$ do not commute with the total spin operators, the linear combination~\eqref{phgen} of their images under the rotations $U_a$ is invariant under global SU(2) transformations. The Hamiltonian~\eqref{phgen} is a non-local $t$-$J$-$V$ like model with additional long-range three-body interaction terms. Specifically, the locally varying single-body potential $\mu_j$, kinetic hopping parameter $t_{jk}$, density-density coupling $V_{jk}$ and spin-exchange coupling $J_{jk}$ are given by
\begin{subequations}\label{coupling1}
	\begin{align}
	\mu_j&=3(m-1)\sum_{k(\neq j)}(w_{jk}-\wb_{jk})+6\sum_{k(\neq j)}|w_{jk}|^2+3\sum_{k\neq l(\neq j)}\wb_{jk}w_{jl},\\
	t_{jk}&=3(m-1)(w_{jk}-\wb_{jk})+6|w_{jk}|^2+3\sum_{l(\neq j,k)}(\wb_{jk}w_{jl}+\wb_{kj}w_{kl}+\wb_{lk}w_{lj}),\\
	\begin{split}V_{jk}&=6(m^2-m-\frac{1}{2})|w_{jk}|^2
	-3(m+\frac{1}{2})\Big[(m-1)(w_{jk}+\wb_{jk})+\sum_{l(\neq j,k)}(\wb_{jk}w_{jl}+\wb_{kl}w_{kj})\Big],
	\end{split}\\
	J_{jk}&=-2(m-1)(w_{jk}+\wb_{jk})+2(2m-1)|w_{jk}|^2 -2\sum_{l(\neq j,k)}(\wb_{jk}w_{jl}+\wb_{kl}w_{kj}),
	\end{align}
\end{subequations}whereas the three-body couplings are $g^{(1)}_{jkl}=(m+\frac{1}{2})(\wb_{jk}w_{jl}+\wb_{kj}w_{kl}+\wb_{lk}w_{lj})$, $g^{(2)}_{jkl}=\wb_{jk}w_{jl}+(2m+1)(\wb_{lj}w_{lk}+\wb_{kl}w_{kj})$, $g^{(3)}_{jkl}=-3(m+\frac{1}{2})(\wb_{jk}w_{jl}+\wb_{kl}w_{kj})$ and $g^{(4)}_{jkl}=-(\wb_{jk}w_{jl}+\wb_{kl}w_{kj})$. On generic lattices the two-body coupling constants are not real such that the model~\eqref{phgen} explicitly breaks time reversal. The parent Hamiltonian~\eqref{phgen} of the infinite MPS is unphysical due to its long-range interaction terms. Nonetheless, it may still be relevant for realistic physical systems provided that it can be deformed into a local Hamiltonian without crossing a phase boundary. In this case, the universal properties of the infinite MPS characterise the ground state of the physical model. For many other CFTs short-range physical models in the same phase as the long-ranged infinite MPS parent Hamiltonians have been found~\cite{Nielsen2013,Nielsen_su2_k,Glasser2015,Tu2014_sun}. We leave the corresponding analysis for the Halperin infinite MPS for future work and instead focus on the low-energy properties of the parent Hamiltonians on one-dimensional chains.

\subsubsection{Parent Hamiltonian on periodic chains}

On a possibly non-uniform periodic chain $\Gamma=\{z_j=e^{i\varphi_j}|\varphi_j\in [0,2\pi)\,\forall j=1,\dotsc,N\}$ the parent Hamiltonian~\eqref{phgen} can be simplified drastically thanks to the existence of numerous discrete Fourier sums. In particular, all three-body terms reduce to two-body terms or can be removed by the addition and subtraction of operators annihilating the infinite MPS. As we show in Appendix~\ref{app:parentHamil}, the infinite MPS for all $m\geq 1$ is an eigenstate of the SU(2) invariant two-body Hamiltonian
\begin{align}
\notag H^{\mathrm{pbc}}&=\frac{\pi^2}{N^2}\Big[\frac{1}{6}\sum_{a=1,2,3}\sum_{j=1}^N\sum_{\sigma=\uparrow,\downarrow}U_a^{\dagger}\big[\big(\Omega'^{\sigma}_j\big)^{\dagger}\Omega'^{\sigma}_j-m\big(C'^{\sigma}_j\big)^{\dagger}C'^{\sigma}_j-(m-1)\big(D'^{\sigma}_j\big)^{\dagger}D'^{\sigma}_j\big]U_a\\
\notag &\quad+\frac{1}{2}(m+1)\sum_{j,k=1}^N\sum_{\sigma=\uparrow,\downarrow}\dda{j\sigma}\an{k\sigma}+\frac{2m}{9}\sum_{a=1,2,3}\sum_{j=1}^N(\Lambda_j^a)^{\dagger}\Lambda_j^a\Big]+E^{\mathrm{pbc}}_0\\
\notag &=\frac{\pi^2}{N^2}\Big[\sum_{j\neq k}\big(\sin\frac{1}{2}(\varphi_j-\varphi_k)\big)^{-2}\big[-n_j+\sum_{\sigma=\uparrow,\downarrow}\dda{j\sigma}\an{k\sigma}+\frac{2m^2+m}{4}n_jn_k+m\,\vec{S}_j\cdot\vec{S}_k\big]\\
\label{phpbc} &\quad+\sum_{j\neq k}w_{jk}(b_j-b_k)[n_j-\frac{2m+1}{4}n_jn_k-\frac{1}{3}\vec{S}_j\cdot\vec{S}_k]+(m-1)\sum_{j\neq k}w_{jk}\sum_{\sigma=\uparrow,\downarrow}\dda{j\sigma}\an{k\sigma}\Big]
\end{align}with energy
\begin{equation}
E_0^{\mathrm{pbc}}=\frac{\pi^2}{N^2}\Big[\frac{1}{6}\big(m+\frac{1}{2}\,\big)^2\,M(M-1)(M-2)-\frac{1}{6}M(N-1)(N-2)
-\big(m^2+\frac{3}{4}+\frac{N}{2}\big)M+\frac{1}{8}(4m^2+4m+1)M^2\Big].
\end{equation}Here, the overall factor $\pi^2/N^2$ ensures that the couplings remain finite in the thermodynamic limit and we introduced $b_j\equiv\sum_{k(\neq j)}w_{kj}$ for $j=1\dotsc,N$. Up to an additional chiral hopping term $\sum_{j\neq k}w_{jk}\sum_{\sigma}\dda{j\sigma}\an{k\sigma}$ that is proportional to $(m-1)$, the Hamiltonian~\eqref{phpbc} defines an extension to non-uniform periodic chains of the inverse-square $t$-$J$-$V$ model~\eqref{hamha} discussed by Ha and Haldane. For a uniform periodic chain with $\varphi_j=2\pi j/N$ one may show that $b_j=0$ and the infinite MPS is an eigenstate of
\begin{equation}\label{phpbc2body}
\frac{\pi^2}{N^2}\Big[\sum_{j\neq k}\sin^{-2}\big(\frac{\pi}{N}(j-k)\big)\Big(\sum_{\sigma}\dda{j\sigma}\an{k\sigma}+\frac{2m^2+m}{4}n_jn_k+m\,\vec{S}_j\cdot\vec{S}_k\Big)+(m-1)\sum_{j\neq k}w_{jk}\sum_{\sigma}\dda{j\sigma}\an{k\sigma}\Big]
\end{equation}with energy
\begin{equation}\label{E0pbcuniform}
\frac{\pi^2}{N^2}\Big[\frac{1}{6}\big(m+\frac{1}{2}\,\big)^2\,M(M-1)(M-2)-\frac{1}{6}M(N-1)(N-2)-(m^2+\frac{3}{4})M+\frac{1}{8}(4m^2-1)M^2\Big].
\end{equation}The Hamiltonian~\eqref{phpbc2body} is exactly equal to Ha and Haldane's model except for the chiral hopping term which vanishes for $m=1$ and for higher $m$ ensures that there is a unique ground state with non-zero momentum, in contrast to the time-reversal invariant model~\eqref{hamha}. Due to the subtraction of the positive terms $\sum_{a,j,\sigma}U_a^{\dagger}\big[\frac{m}{6}(C'^{\sigma}_j)^{\dagger}C'^{\sigma}_j+\frac{m-1}{6}(D'^{\sigma}_j)^{\dagger}D'^{\sigma}_j\big]U_a$ in~\eqref{phpbc} we cannot prove rigorously that~\eqref{phpbc2body} is bounded below by the energy~\eqref{E0pbcuniform}. However, for $m=1$ it is known from other work~\cite{ha1992,Kawakami1992,Schuricht2006,Tu2014_sun} that the infinite MPS is indeed the exact ground state of~\eqref{phpbc2body}, which is just the SU(3) HS model. Exact diagonalisation in small systems shows that the infinite MPS is the exact ground state of~\eqref{phpbc2body} also for $m=2,3$ and we expect that this persists in the thermodynamic limit and for higher values of $m$.

\subsubsection{Parent Hamiltonian on open chains}
The open chain $\Gamma=\{u_j=\cos\theta_j|\theta_j\in[0,\pi]\,\forall j=1,\dotsc,N\}$ can be understood as the projection onto the real axis of the periodic chain $\tilde{\Gamma}=\{z_j=e^{i\theta_j},z_{\bar{j}}=e^{-i\theta_j}|j=1,\dotsc,N\}$. In order to make use of the Fourier sum identities for periodic chains, we construct the parent Hamiltonian on open chains from the operators $\Phi''_j=-(z_j-\zb_j)\Phi_j=-\sum_{\lambda}\sum_{k(\neq j)}((z_j-\zb_j)/(u_k-u_j))\mathcal{Q}^{\lambda}_j\mathcal{O}^{\lambda}_k=\sum_{\lambda}\sum_{k(\neq j)}(w_{jk}+w_{j\kb})\mathcal{Q}^{\lambda}_j\mathcal{O}^{\lambda}_k$ that depend on the angles $\theta_j$ through the functions $w_{ij}=(z_i+z_j)/(z_i-z_j)$ and $w_{i\jb}=(z_i+\zb_j)/(z_i-\zb_j)$. The operators annihilating the Halperin infinite MPS on open chains derived in this way from the null fields~\eqref{nullfieldspin} and~\eqref{nullfieldcharge} are summarised in the third column of Tab.~\ref{tab:oppbc}. In Appendix~\ref{app:parentHamil} we show that for all $m\geq 1$ the infinite MPS on open chains is an eigenstate of the SU(2) invariant two-body Hamiltonian
\begin{align}
\notag H^{\mathrm{obc}}&=\frac{\pi^2}{N^2}\Big[\frac{1}{6}\sum_{a=1,2,3}\sum_{j=1}^N\sum_{\sigma=\uparrow,\downarrow}U_a^{\dagger}\big[(\Omega''^{\sigma}_j)^{\dagger}\Omega''^{\sigma}_j-m(C''^{\sigma}_j)^{\dagger}C''^{\sigma}_j-(m-1)(D''^{\sigma}_j)^{\dagger}D''^{\sigma}_j\big]U_a\\
\notag &+2m\sum_{j,k=1}^N\sum_{\sigma=\uparrow,\downarrow}\dda{j\sigma}\an{k\sigma}+\frac{2m}{9}m_{a=1,2,3}\sum_{j=1}^N(\Lambda''^a_j)^{\dagger}\Lambda''^a_j\Big]+E^{\mathrm{obc}}_0\\
\notag
&=\frac{\pi^2}{N^2}\Big[-\sum_{j\neq k}\left[w_{jk}(c_j-c_k)+w_{j\bar{k}}(c_j+c_k)\right]\big[\frac{2m+1}{4} n_j n_k+\frac{1}{3}\vec{S}_j\cdot\vec{S}_k-n_j\big]\\
\label{phobc}&+\sum_{j\neq k}\big[\sin^{-2}\frac{\theta_j-\theta_k}{2}+\sin^{-2}\frac{\theta_j+\theta_k}{2}\big]\big[\frac{2m^2+m}{4} n_j n_k+m \vec{S}_j\cdot \vec{S}_k+\sum_{\sigma=\uparrow,\downarrow}\dda{j\sigma}\an{k\sigma} -n_j\big]\Big]
\end{align}with energy
\begin{multline}
E_0^{\mathrm{obc}}=\frac{\pi^2}{N^2}\Big[\frac{2}{3}\Big(m+\frac{1}{2}\Big)^2M(M-1)(M-2)+\Big(m^2+\frac{7}{2}m+\frac{3}{2}\Big)M(M-1)\\
-\frac{1}{2}(2m+1)M(M-2)-2(N-1)NM-2(N-2)M-\frac{7}{2}M-\frac{3m}{2}M\Big],
\end{multline}where $c_j\equiv w_{\jb j}+\sum_{k(\neq j)}(w_{kj}+w_{\kb j})$. The Hamiltonian~\eqref{phobc} is a time-reversal invariant generalisation of Ha and Haldane's inverse-square $t$-$J$-$V$ model~\eqref{hamha} to non-uniform one-dimensional chains with open boundary conditions. On a type-I uniform open chain we have $w_{jk}(c_j-c_k)+w_{j\bar{k}}(c_j+c_k)=0$~\cite{Tu2015obc} such that the Hamiltonian~\eqref{phobc} simplifies to
\begin{equation}\label{phobc2body}
\frac{\pi^2}{N^2}\sum_{j\neq k}\Big[\sin^{-2}\frac{\pi(j-k)}{2N}+\sin^{-2}\frac{\pi(j+k-1)}{2N}\Big]
\Big[\frac{2m^2+m}{4} n_j n_k+m \,\vec{S}_j\cdot \vec{S}_k+\sum_{\sigma}\dda{j\sigma}\an{k\sigma} -n_j\Big].
\end{equation}The relative strength of the hopping parameter, density-density interaction and spin exchange in the open-boundary parent Hamiltonian~\eqref{phobc2body} are the same as in the periodic model~\eqref{hamha} studied by Ha and Haldane. However, the coupling constants $t_{jk}$, $V_{jk}$ and $J_{jk}$ are proportional not to just the inverse square $|z_j-z_k|^{-2}$ of the chord distance between $z_j$ and $z_k$, but instead to the sum $|z_j-z_k|^{-2}+|z_j-\zb_k|^{-2}$ of the inverse square chord distances between $z_j$ and $z_k$ as well as $z_j$ and the mirror image $\zb_k$. This is akin to the modification of the inverse-square coupling strength in the HS model due to open boundaries~\cite{Bernard1995,Simons1994}. Similarly to the periodic case we are at the present time unaware of any way to show analytically that the infinite MPS is the ground state of the Hamiltonian~\eqref{phobc2body} for $m\geq 2$ due to the subtraction of positive operators in~\eqref{phobc}. However, for $m=1$ we know that this is the case thanks to the connection to the SU(3) open-boundary HS model~\cite{Tu2015obc} and we have confirmed by exact diagonalisation that~\eqref{phobc2body} is bounded below by $E_0^{\mathrm{obc}}$ for $m=2,3$.

\subsection{Inverse-square $t$-$J$-$V$ models as two-component Luttinger liquids\label{sec:LLCFT}}

\subsubsection{Rapidity description for low-energy spectrum of periodic and open models\label{sec:Rapidities}}

\begin{figure}[t]
	\centering
	\includegraphics[width=\textwidth]{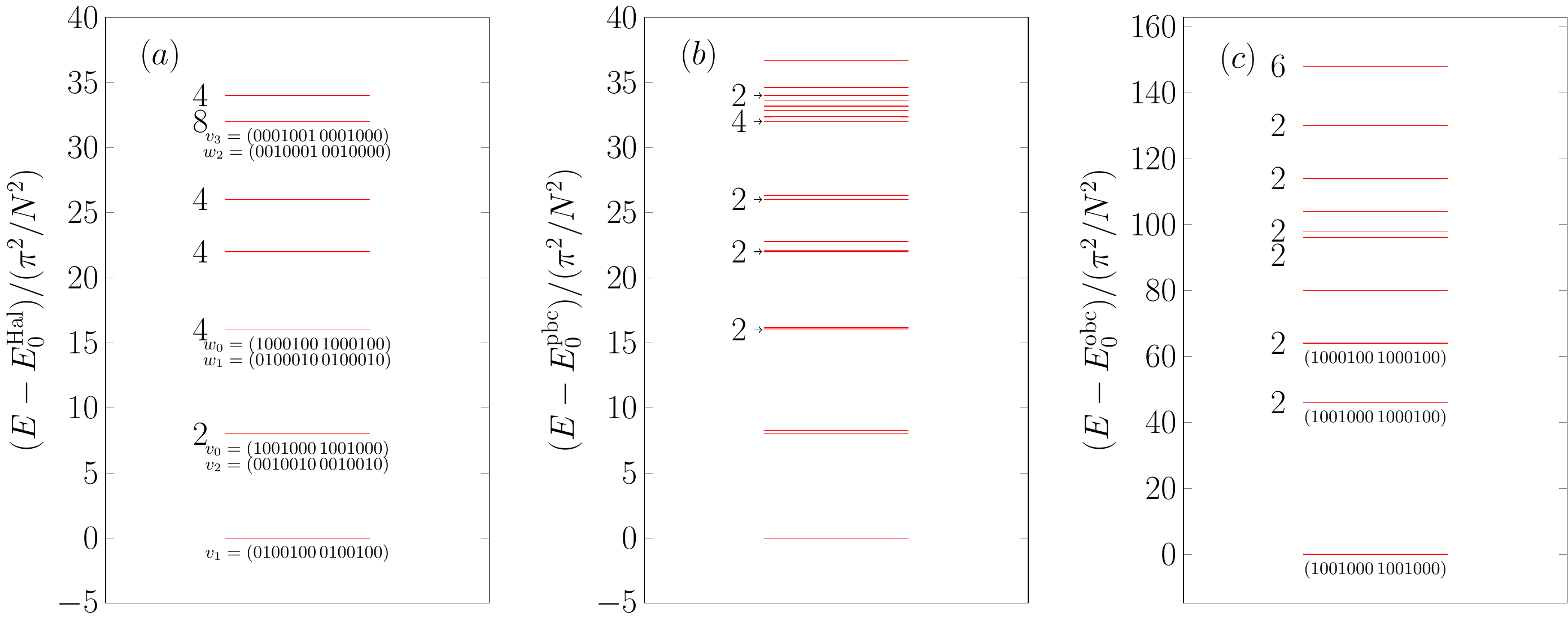}
	\caption{Lowest eigenvalues and their degeneracy of the models discussed in the main text on a chain with $N=14$ sites at filling fraction $2/7$ corresponding to $m=3$ as obtained from exact diagonalisation. $(a)$ Spectrum of Ha and Haldane's inverse-square $t$-$J$-$V$ model~\eqref{hamha} including the assignment of the rapidity sets $v_k$ and $w_l$ corresponding to primary fields. Due to finite-size effects the energy of $v_3$ and $w_2$ is greater than that of some descendant states. $(b)$: Spectrum of the infinite MPS parent Hamiltonian~\eqref{phpbc2body} on periodic chains. $(c)$ Spectrum of the infinite MPS parent Hamiltonian~\eqref{phobc2body} on open chains including the rapidity sets corresponding to the three lowest levels. In units of $\pi^2/N^2$ the ground state energies are $E^{\mathrm{Hal}}_0=-16$, $E^{\mathrm{pbc}}_0=-32$, $E^{\mathrm{obc}}_0=-1164$.}
	\label{fig:SpectraED}
\end{figure}

Exact diagonalisation of the $t$-$J$-$V$ models~\eqref{hamha}, \eqref{phpbc2body} and~\eqref{phobc2body} on periodic and open chains shows that the spectrum of all three Hamiltonians contains many eigenvalues which are rational in units of $\pi^2/N^2$ (see Fig.~\ref{fig:SpectraED} for the low-energy spectra on a chain with $N=14$ sites and filling fraction $\nu=2/7$). For Ha and Haldane's periodic Hamiltonian~\eqref{hamha} and the parent Hamiltonian~\eqref{phobc2body} on type-I uniform chains, the number of rational eigenvalues that are lower in energy than the first irrational eigenvalue increases with growing system size $N$ such that in the thermodynamic limit we expect the entire low-energy spectrum to consist of rational eigenvalues. We observed that the low-lying rational eigenvalues of these two models at filling fraction $\nu=2/(2m+1)$ and vanishing total spin $S^3_{\mathrm{tot}}$ are described by rapidity sets obeying the same generalised Pauli exclusion principle. A rapidity set $v\equiv\{m_1,m_2,\dotsc\}$ for a system of size $N$ is a collection of non-identical integers $m_1<m_2<\dotsb$ in the range $m_i\in[1,\dotsc,N]$. The rapidity set $v$ may be represented by the corresponding occupation number sequence $(n_1,\dotsc,n_N)$ with $n_j\in\{0,1\}$ and $n_j=1$ ($n_j=0$) if there is (not) a rapidity $m_i$ in $v$ such that $m_i=j$. For periodic and open boundary conditions, we assign to the rapidity set $v$ the energy
\begin{subequations}\label{enrap}
\begin{gather}\label{enrappbc}
E^{\mathrm{pbc}}_{v}=2\frac{\pi^2}{N^2}\sum_im_i(m_i-N)+\tilde{E}^{\mathrm{pbc}}(m,N),\\
\label{enrapobc}
E^{\mathrm{obc}}_v=2\frac{\pi^2}{N^2}\sum_{i}(m_i^2-N^2),
\end{gather}
\end{subequations}where $\tilde{E}^{\mathrm{pbc}}(m,N)=2N/(3(2m+1)^3)(32m^3+4m^2(N^2-6N-1)+4 m (2N^2 + 3 N-7)-3 (N^2-4N+3))$ is a term that depends only on $N$ and $m$. Up to an overall factor of $2$, the rapidity dispersion relations are hence the same as for the periodic and open HS model~\cite{Haldane1992,Bernard1995,Tu2015obc}. For periodic boundary conditions the total lattice momentum associated to the rapidity sequence $v$ is proportional to the sum of the rapidities
\begin{equation}\label{momrap}
P_v=\frac{2\pi}{N}\Big[\sum_im_i \mod N \Big]
\end{equation}in complete analogy to the periodic HS model~\cite{Haldane1992}. In the sector of vanishing total spin $S^3_{\mathrm{tot}}$ and filling fraction $\nu=2/(2m+1)$ the distinct energy and momentum eigenvalues of Ha and Haldane's inverse-square $t$-$J$-$V$ model~\eqref{hamha} and its open-boundary generalisation~\eqref{phobc2body} correspond precisely to those obtained from all rapidity sets obeying the following generalised Pauli principle: Firstly, between any two occupied rapidity orbitals there must lie at least $m-1$ empty orbitals such that $m_{i+1}-m_{i}\geq m$, and secondly, out of any $2m+1$ successive orbitals at most two can be occupied, i.e. $m_{i+2}-m_{i}\geq 2m+1$. As expected, for $m=1$ this reduces to the well-known generalised Pauli principle $m_{i+2}-m_i\geq 3$ characterising the SU(3) HS model~\cite{Haldane1992}.

In the thermodynamic limit, the lowest-lying states in Ha and Haldane's periodic model~\eqref{hamha} are associated with the rapidity sets 
\begin{subequations}\label{OccNumbSequence}
\begin{gather}
v_k=(0^k \,1 \,0^{m-1}\, 1\, 0^{m} \,1 \,0^{m-1}\, 1\, 0^{m}\dotsm 1\, 0^{m-1} \, 1\, 0^{m-k}),\\
w_l=(0^l \,1 \,0^{m}\, 1\, 0^{m-1} \,1 \,0^{m}\, 1\, 0^{m-1}\dotsm 1\, 0^{m} \, 1\, 0^{m-1-l}),
\end{gather}
\end{subequations}where the symbol $0^i$ indicates $i$ successive entries that are equal to zero and we choose $k=0,\dotsc,m$ and $l=0,\dotsc,m-1$. These rapidity sets have energy and momentum eigenvalues
\begin{subequations}\label{enlowrap}
	\begin{gather}\label{env}
	E_{v_k}=E_{v_0}+\frac{2\pi^2}{N}\frac{2}{2m+1}[k^2-(m-1)k],\\
	\label{enw}
	E_{w_l}=E_{v_0}+\frac{2\pi^2}{N}\frac{2}{2m+1}[1+l^2-(m-2)l],\\
	P_{v_k}=2k_F (m+2+2k),\\
	P_{w_l}=2k_F(m+2+2l+1).
	\end{gather}
\end{subequations}Hence, $v_k$ ($w_l$) describes the configuration of lowest energy in which $2k$ ($2l+1$) hard-core particles were shifted from the right branch to the left branch of the single-particle dispersion relation compared to $v_0$. They correspond precisely to the low-energy eigenstates constructed by Ha and Haldane for the model~\eqref{hamha} in terms of their Jastrow wave functions~\cite{ha1992}. In particular, the ground state for a bosonic system with $m$ odd is described by the rapidity set $v_{(m-1)/2}$, whereas in the fermionic case the two rapidity sets $v_{m/2-1}$ and $v_{m/2}$ have the same lowest energy. Gapless excitations derived from the low-energy states $v_k$ and $w_l$ are associated with shifts of single rapidities at the edges of the sequence. Fig.~\ref{fig:SpectraED}~(a) illustrates the assignment of the rapidity sequences $v_l$ and $w_k$ to the low-lying levels in the spectrum of Ha and Haldane's model at $m=3$ on a chain with $N=14$ sites.\par

Up to a shift of the ground state energy the low-energy spectrum of the infinite MPS parent Hamiltonian~\eqref{phpbc2body} is similar to that of Ha and Haldane's model (see Fig.~\ref{fig:SpectraED}~(b)). However, the perfect degeneracy of many excited states is lifted, leading to the appearance of low-lying irrational eigenvalues that cannot be described using rapidity sets.\par  

The ground state of the open-boundary model~\eqref{phobc2body} is associated for all $m$ with the rapidity sequence $v_0^{\mathrm{obc}}=(1 \, 0^{m-1} \, 1 \, 0^m \, 1\, 0^{m-1} \, 1 \, 0^m \, \dotsm 1 \, 0^{m-1} \, 1 \, 0^m )$. The low-lying excitations are obtained by a finite number of shifts of single rapidities to the right by one orbital compared to $v_0^{\mathrm{obc}}$. If $k$ such shifts have been performed, the excitation energy scales as $4k\pi^2/N+\mathcal{O}(N^{-2})$ such that all these excitations are gapless.\par

The generalised Pauli principle proposed above is identical to a $(k,r)$ admissibility condition of the kind proposed in Ref.~\onlinecite{Estienne2012} for $S_{M/2}\otimes S_{M/2}$ symmetric Jack polynomials with $k=1$ and $r=m+1$ and when neglecting the spin dressing of partitions. It is known that the Jack polynomial eigenstates of the spin-less Calogero-Sutherland model~\cite{Sutherland1971a,Sutherland1971b,Calogero1969} are also eigenstates of the HS model with rational eigenvalues (see for instance Ref.~\onlinecite{Kuramoto2009}). Based on the similarities between the generalised Pauli principle described above and the $(1,m+1)$ admissibility condition one may thus conjecture that the $S_{M/2}\otimes S_{M/2}$ symmetric Jack polynomial eigenstates of the spinful Calogero-Sutherland model~\cite{Kuramoto2009} are also eigenstates of Ha and Haldane's inverse-square $t$-$J$-$V$ model~\eqref{hamha}. In addition we expect that the excited state wave functions of the infinite MPS parent Hamiltonian can be obtained by the insertion into the CFT correlator of additional CFT fields evaluated at $0$ and $\infty$~\cite{Herwerth2015}.

\subsubsection{Determination of scaling dimensions for the periodic model}

The low-energy physics of the quantum critical inverse-square $t$-$J$-$V$ model~\eqref{hamha} on a periodic chain with $N$ sites is described by a continuum CFT on a (1+1)-dimensional space-time cylinder with periodic boundary conditions and length $Na$ in the spatial direction. The Hilbert space of this CFT consists of states at left- and right-moving Virasoro levels $n,\nb\in\mathbb{N}$ which are descended from primary states with chiral and anti-chiral conformal dimensions $h,\bar{h}$. In units where the lattice spacing is $a=1$ and $\hbar=1$, the energy and momentum of these states is given by
\begin{subequations}\label{cftpred}
\begin{gather}
\label{cftenergy}
E(h,\hb,n,\nb)=\epsilon^{\infty}N-\frac{ u\pi c}{6N}+\frac{ 2\pi u}{N}(h+\hb+n+\nb),\\
\label{cftmom}
P(h,\hb,n,\nb)=\frac{2\pi}{N}(h-\hb+n-\nb).
\end{gather}
\end{subequations}Here, $u$ is the characteristic velocity of the system, $c$ is the central charge and $\epsilon^{\infty}$ is the average ground state energy per unit length in the thermodynamic limit. Since the primary states in the Luttinger CFT of~\eqref{hamha} are associated with the rapidity sets $v_k$ and $w_l$ we can compare the exact expressions~\eqref{enlowrap} for their energy and momentum with the CFT predictions~\eqref{cftpred} to extract their conformal dimensions $h=\bar{h}$. Single-particle excitations above the primary states with momentum difference $\Delta P=\pm2\pi/N$ correspond to descendants at Virasoro level $n=1,\bar{n}=0$ ($n=0,\bar{n}=1$) and are described by shifts of single rapidities at the right (left) end of the sequence towards the right (left). Since all these states have excitation energies $\Delta E=2\pi^2/N+\mathcal{O}(1/N^2)$, the low-energy effective theory of the Hamiltonian~\eqref{hamha} has a single characteristic velocity $u=\pi$; in particular there is no spin-charge separation.\par  

For bosonic systems with odd values of $m$ we identify the identity Verma module $h=\bar{h}=0$ with the non-degenerate ground state $v_{(m-1)/2}$. Then, the rapidity sets $v_{k\equiv(m-1)/2+\tilde{k}}$ and $w_{l\equiv (m-2)/2+\tilde{l}}$ correspond to primary fields of conformal dimension
\begin{subequations}\label{scaling1}
	\begin{gather}
	h_{\tilde{k}}=\frac{(2\tilde{k})^2}{4(2m+1)},\\
	h_{\tilde{l}}=\frac{1}{4}+\frac{(2\tilde{l})^2}{4(2m+1)},
	\end{gather}
\end{subequations}where $\tilde{k}\in \{-(m-1)/2,\dotsc,(m-1)/2+1\}$ and $\tilde{l}\in\{(m-2)/2,\dotsc,(m-2)/2+1\}$ run in integer steps. For the SU(3) HS model at $m=1$ this gives three different primary states with conformal dimensions $h_{\tilde{k}=0}=0$ and $h_{\tilde{k}=1}=1/3=h_{\tilde{l}=0}$ as expected from $\mathfrak{su}(3)_1$. Note that this procedure cannot be used to extract the central charge of the low-energy effective CFT for the inverse-square $t$-$J$-$V$ model since the Hamiltonian~\eqref{hamha} contains long-range interactions and the scaling of the ground state energy in critical non-local models generally violates the CFT prediction~\eqref{cftenergy}~\cite{Tu2014_sun,Bondesan2014}.\par  
For the fermionic models with even $m$ the identification of the identity module corresponding to $h=\bar{h}=0$ is not straightforward due to the double degeneracy of the ground state of Ha and Haldane's model~\eqref{hamha}. Indeed, the naive assignment of the identity module to either $v_{m/2-1}$ or $v_{m/2}$ gives incorrect results since the predicted list of conformal dimensions does not include the value $h=(m+1)/(2(2m+1))$ observed in the spin correlation function. Instead, we suggest to enlarge the set of states by considering also sets of rapidities with half-integer values $m_i\in[3/2,\dotsc,N+1/2]$ but the same generalised exclusion principle and dispersion relation as described above for integer-valued rapidity sets. This leads to the appearance of $2m+1$ additional low-energy states corresponding to the occupation number sequences~\eqref{OccNumbSequence} and with energies and momenta given by the expressions on the LHS of~\eqref{enlowrap} after the replacement $k\mapsto k+1/2$ and $l\mapsto l+1/2$. The collection of low-energy states for both integer and half-integer rapidities contains a unique configuration of lowest energy that is associated with the half-integer rapidity set described by the occupation number sequence $v_{m/2-1}$. After the identification of this state with the identity Verma module $h=\bar{h}=0$, the enlarged set of low-energy states corresponds to primary fields with conformal dimensions given by~\eqref{scaling1}, where $\tilde{k}\in \{-(m-1)/2,\dotsc,(m-1)/2+3/2\}$ and $\tilde{l}\in\{(m-2)/2,\dotsc,(m-2)/2+3/2\}$ now run in half-integer steps. In particular, the conformal dimension $h_{\tilde{l}=\pm1/2}=(m+1)/(2(2m+1))$ agrees with the critical exponent observed in the spin correlation function. Since the rapidities $m_i$ correspond to spinon quasi-momenta we expect that half-integer rapidity sets describe a system that is coupled to an external gauge field of flux $\phi=1/2$, or equivalently, subject to anti-periodic boundary conditions for the fermionic particles~\cite{Fukui1996,Liu1997}. Therefore, the addition of states associated to half-integer rapidity sets is motivated by... Following Ref.~\onlinecite{Liu1997}, we have attempted to generalise Ha and Haldane's Hamiltonian to systems with anti-periodic boundary conditions. However, the resulting Hamiltonian does not have a lower ground state energy than~\eqref{hamha} such that its rational eigenvalues are not described by the half-integer rapidity sets introduced above.

\subsubsection{Action description for low-energy effective CFT of bosonic periodic model\label{sec:action}}
Since the CFT that describes the low-energy physics of the periodic inverse-square $t$-$J$-$V$ model~\eqref{hamha} has central charge $c=2$ we expect it to be a theory of a two-component massless free boson $X=(X^1,X^2)$ compactified on a two-dimensional torus. On a two-dimensional word-sheet parametrised by Euclidean coordinates $x_{\mu}$ with $\mu=0,1$, the most general such theory is described by the Euclidean action~\cite{Sule2015}
\begin{equation}
S_E=\frac{1}{4\pi}\int dx_0dx_1[G_{ab}\partial_{\mu}X^a\partial_{\mu}X^b+iB_{ab}\epsilon_{\mu\nu}\partial_{\mu}X^a\partial_{\nu}X^b],
\end{equation}where $G_{ab}$ and $B_{ab}$ are real symmetric and anti-symmetric matrices, respectively. In the case without orbifolding when both bosonic fields obey periodic boundary conditions the partition function of this theory on a world-sheet torus can be evaluated explicitly and yields the spectrum of scaling dimensions~\cite{Sule2015}
\begin{equation}\label{spectrum}
\Delta\big(\mv{n},\mv{w},\{N_{n_L}^L\},\{N_{n_R}^R\}\big)=\mv{p}_L^T \cdot G \cdot \mv{p}_L + \mv{p}_R^T \cdot G \cdot \mv{p}_R +\sum_{n_L > 0} n_L N_{n_L}^L + \sum_{n_R > 0} n_R N_{n_R}^R
\end{equation}where
\begin{equation}
\mv{p}_{L,R}=\frac{1}{2}[G^{-1}(\mv{n}-B\cdot\mv{w})\pm\mv{w}].
\end{equation}Here, $\mv{n}=(n_1,n_2)^T$ and $\mv{w}=(w_1,w_2)^T$ are the winding numbers of the two-component boson field around the two non-contractible loops of the torus and the collection of integer numbers $\{N_{n_L}^L\},\{N_{n_R}^R\}$ specifies the descendant level. When $m$ is odd the low-lying scaling dimensions in the spectrum~\eqref{spectrum} for the choice $G_{ab}=\big(\begin{smallmatrix}
m+1 & m\\
m & m+1\\
\end{smallmatrix}\big)/2$ and $B_{ab}=\big(\begin{smallmatrix}
	0 & 1\\
	-1 & 0\\
\end{smallmatrix}\big)/2$ correspond precisely to the conformal dimensions~\eqref{scaling1} that we identified from the finite-size scaling of the lowest eigenvalues of the model~\eqref{hamha}. As expected, for $m=1$ the matrix $G$ is proportional to the inverse of the Cartan matrix of $\mathfrak{su}(3)$~\cite{Sule2015}. This completes the identification of the low-energy effective CFT for Ha and Haldane's periodic model~\eqref{hamha} in the bosonic case. For the fermionic models at even $m$,  we were not able to reproduce the observed conformal dimensions~\eqref{scaling2} by making appropriate choices for $G$ and $B$ in~\eqref{spectrum}. This may indicate that the low-energy effective theory of the fermionic systems is a more general orbifolded two-component free boson CFT. We mention that the fermionic model with $m=2$ may be related to certain $\mathcal{N}=4$ superconformal field theories~\cite{Fokkema2016}.

\section{Conclusion}\label{sec:conclusion}

Starting from deformations of the CFT $\mathfrak{su}(3)_1$ we proposed a series of many-body states parametrised by a natural number $m$ that describe systems of interacting spin one-half hard-core bosons (fermions) for odd (even) $m$ and whose wave functions have a Jastrow part identical to that of the $(m+1,m+1,m)$ Halperin FQH state. We derived SU(2) invariant parent Hamiltonians for these states on arbitrary one- and two-dimensional lattices. On two-dimensional lattices the wave function corresponds precisely to the $(m+1,m+1,m)$ Halperin state with the positions of the particles restricted to the lattice sites, while the parent Hamiltonian is a long-range chiral $t$-$J$-$V$ model with additional three-body interaction terms which is expected to possess abelian anyonic excitations in analogy with the continuum system. On one-dimensional chains with periodic (open) boundary conditions the parent Hamiltonian contains only two-body terms and for $m=1$ reduces to the periodic (open) SU(3) HS model. We were thus able to generalise a periodic inverse-square $t$-$J$-$V$ model proposed and studied in Ref.~\onlinecite{ha1992} to chains with open boundary conditions, whereas the parent Hamiltonian on periodic chains agrees with the former model up to an additional chiral hopping term. The distinct low-lying eigenvalues in the finite-size spectrum of the time-reversal invariant periodic inverse-square $t$-$J$-$V$ model and its open-boundary generalisation are rational and can be described by rapidity sets with the same generalised Pauli exclusion principle. We extracted the conformal dimensions of several primary fields in the low-energy effective CFT of the periodic model and for odd $m$ identified a two-component compactified free-boson theory with the same spectrum of scaling dimensions.\par
There are several interesting questions that could be addressed in future work. Firstly, it may be possible to truncate the long-range interactions in the parent Hamiltonian on two-dimensional lattices without crossing a phase boundary. This would provide a short-range Hamiltonian with few-body interaction terms that stabilises a lattice analogue of the Halperin state with abelian anyonic excitations and which may be experimentally realisable. Secondly, it is known that in continuous two-dimensional systems deformations of the CFT $\mathfrak{su}(3)_k$ at levels $k\geq 2$ lead to spin-singlet FQH states with non-abelian anyonic excitations~\cite{Ardonne1999}. It would be interesting to use the infinite MPS construction to define the lattice analogues of these non-abelian states and study their properties in one and two dimensions.

\section{Acknowledgment}

The authors acknowledge discussions with Ignacio Cirac, Germ\'an Sierra, Anne Nielsen and Ying-Hai Wu. This work has been supported by the EU project SIQS. AH acknowledges funding by the European Research Council (ERC) grant WASCOSYS (No. 636201).

\appendix
\section{Vertex operators of a chiral free boson\label{app:cft}}
After a compactification of the target space to a circle of finite radius, the massless real free boson field splits into decoupled chiral and anti-chiral parts $\phi_L(z),\phi_R(\zb)$, where $z,\zb\in\mathbb{C}$ are the coordinates on the complex plane~\cite{Schellekens_CFT}. The primary fields in the chiral sector consist of the left-moving U(1) current $J=i\partial\phi_L$ and the chiral vertex operators $V_{\alpha}^L=\nor{e^{i\alpha\phi_L}}$, where $:\dotsb:$ denotes normal ordering. The OPE of two vertex operators is given by~\cite{DiFrancesco}
\begin{equation}\label{opechiralvertex}
V_{\alpha}^L(z)V_{\beta}^L(w)=(z-w)^{\alpha\beta}V^L_{\alpha+\beta}(w)+\alpha\times(z-w)^{\alpha\beta+1}\nor{J(w)V^L_{\alpha+\beta}(w)}+\mathcal{O}\big((z-w)^{\alpha\beta+2}\big).
\end{equation}Correspondingly, the vacuum expectation value of a product of $N$ chiral vertex operators takes the form~\cite{DiFrancesco}
\begin{equation}\label{corrvertexop}
\ex{V_{\alpha_1}^L(z_1)\dotsm V_{\alpha_N}^L(z_N)}=\prod_{1 \leq i<j\leq N}(z_i-z_j)^{\alpha_i\alpha_j}\times\begin{cases}
1 \quad &\mbox{if }\sum_{i=1}^{N}\alpha_i=0\\
0 &\mbox{otherwise}
\end{cases},
\end{equation}where the constraint $\sum_{i=1}^{N}\alpha_i=0$ is a consequence of the global U(1) symmetry of the free boson theory. The correlation function of the U(1) current with a product of chiral vertex operators is given by
\begin{equation}\label{corru1wardid}
\ex{J(z)V_{\alpha_1}^L(z_1)\dotsm V_{\alpha_N}^L(z_N)}=\sum_{k=1}^N\frac{\alpha_k}{z-z_k}\times \ex{V_{\alpha_1}^L(z_1)\dotsm V_{\alpha_N}^L(z_N)}.
\end{equation}

\section{Parent Hamiltonian on periodic and open chains\label{app:parentHamil}}
In this appendix, we prove that at filling fraction $\nu=2/(2m+1)$ and vanishing total spin $S^3_{\mathrm{tot}}=0$ the parent Hamiltonian of the Halperin infinite MPS on periodic and open chains is given by a two-body operator as claimed in~\eqref{phpbc} and~\eqref{phobc}. Let us consider a non-uniform periodic chain $\Gamma=\{z_j=e^{i\varphi_j}|\varphi_j\in [0,2\pi)\,\forall j=1,\dotsc,N\}$ such that the infinite MPS is annihilated by the operators in the second column of Tab.~\ref{tab:oppbc}. The positive operator
\begin{align}\label{phgen_step1}
\notag&\sum_{j=1}^N\sum_{\sigma}\big[\big(\Omega'^{\sigma}_j)^{\dagger}\Omega'^{\sigma}_j+\big(C'^{\sigma}_j)^{\dagger}C'^{\sigma}_j+\big(D'^{\sigma}_j)^{\dagger}D'^{\sigma}_j\big]=-(m-1)\sum_{j\neq k}(w_{jk}+\wb_{jk})\big[(m+\frac{1}{2})n_jn_k-n_j+2S^3_jS^3_k\big]+\\
\notag
& (m-1)\sum_{j\neq k,\sigma}(w_{jk}-\wb_{jk})\dda{j\sigma}\an{k\sigma}+\sum_{j\neq k}|w_{jk}|^2\big[2n_j+2\sum_{\sigma}\dda{j\sigma}\an{k\sigma}+(m^2-m-\frac{1}{2})n_jn_k+2(2m-1)S^3_jS^3_k\big]+\\
\notag& \msum_{j,k,l}(\wb_{jk}w_{jl}+\wb_{kj}w_{kl}+\wb_{lk}w_{lj})\big[\sum_{\sigma}\dda{k\sigma}\an{l\sigma}+\frac{1}{3}(m+\frac{1}{2})^2n_jn_kn_l\big]+\msum_{j,k,l}\wb_{jk}w_{jl}\big[n_j+n_jS^3_kS^3_l]+(m-1)^2\ntot\\
& +\msum_{j,k,l}(\wb_{jk}w_{jl}+\wb_{kl}w_{kj})\big[-(m+\frac{1}{2})n_jn_k-2S^3_jS^3_k+(2m+1)S^3_jS^3_kn_l-\sum_{\sigma}\dda{k\sigma}\an{j\sigma}\big((m+\frac{1}{2})n_l+s_\sigma S^3_l\big)\big]
\end{align}can be simplified by noting that the complex numbers $w_{jk}$ are purely imaginary such that $w_{jk}-\wb_{jk}=2w_{jk}$ and the terms proportional to $w_{jk}+\wb_{jk}$ vanish. Since $w_{12}w_{13}+w_{21}w_{23}+w_{31}w_{32}=1$ for any three pairwise different complex numbers $z_1,z_2,z_3\in\mathbb{C}$ we have
\begin{gather}
\sideset{}{'}\sum_{j,k,l}(w_{jk}w_{jl}+w_{kj}w_{kl}+w_{lj}w_{lk})n_jn_kn_l=\msum_{j,k,l}n_jn_kn_l=(\ntot-2)(\ntot-1)\ntot,\\
\msum_{\substack{j,k,l\\\sigma}}(w_{jk}w_{jl}+w_{kj}w_{kl}+w_{lj}w_{lk})\dda{k\sigma}\an{l\sigma}=\msum_{\substack{j,k,l\\\sigma}}\dda{k\sigma}\an{l\sigma}=(N-2)\Big[\sum_{\sigma}(Y^{\sigma})^{\dagger}Y^{\sigma}-n^{\mathrm{tot}}\Big],\\
\msum_{j,k,l}(w_{jk}w_{jl}+w_{kj}w_{kl})S^3_jS^3_kn_l=\msum_{j,k,l}(1-w_{lj}w_{lk})S^3_jS^3_kn_l=\big((S^3_{\mathrm{tot}})^2-\frac{1}{4}\ntot\big)(\ntot-2)-\msum_{j,k,l}w_{jk}w_{jl}n_jS^3_kS^3_l
\end{gather}as well as
\begin{align}\label{CDcalc}
\notag &\msum_{j,k,l,\sigma}(w_{jk}w_{jk}+w_{kj}w_{kl})\dda{k\sigma}\an{j\sigma}\big((m+\frac{1}{2}\,)n_l+s_{\sigma} S^3_l\big)=(m+1)\sum_{j,\sigma}\big(C'^{\sigma}_j)^{\dagger}C'^{\sigma}_j+m\sum_{j,\sigma}\big(D'^{\sigma}_j)^{\dagger}D'^{\sigma}_j\\
&+\big[\sum_{\sigma}(Y^{\sigma})^{\dagger}Y^{\sigma}-\ntot\big]\Big[(m+\frac{1}{2}\,)\left(\ntot-1\right)+\big(S^3_{\mathrm{tot}}-\frac{1}{2}\big)\Big]+\sum_{j\neq k}w_{jk}^2[(m+\frac{1}{2}\,)n_jn_k+2S^3_jS^3_k],
\end{align}where we introduced the operators $Y^{\sigma}=\sum_j\an{j\sigma}$ that annihilate the infinite MPS as shown in Sec.~\ref{sec:iMPSSymmProp}. For any collection $z_1,\dotsc,z_N$ of pairwise different complex numbers of unit absolute value one finds $\sum_{k(\neq i,j)}w_{ki}w_{kj}=N-2+2w_{ij}^2+w_{ij}(b_i-b_j)$ with $b_i\equiv\sum_{j(\neq i)}w_{ji}$. This implies
\begin{gather}
\msum_{j,k,l}(w_{jk}w_{jl}+w_{kj}w_{kl})n_jn_k=\msum_{j,k,l}(1-w_{lj}w_{lk})n_jn_k=-\sum_{j\neq k}[2w_{jk}^2+w_{jk}(b_j-b_k)]n_jn_k,\\
\msum_{j,k,l}(w_{jk}w_{jl}+w_{kj}w_{kl})S^3_jS^3_k=\msum_{j,k,l}(1-w_{lj}w_{lk})S^3_jS^3_k=-\sum_{j\neq k}[2w_{jk}^2+w_{jk}(b_j-b_k)]S^3_jS^3_k,\\
-\msum_{j,k,l}w_{jk}w_{jl}n_j=(N-1)(N-2)\ntot+\sum_{j\neq k}[4w_{jk}^2+2w_{jk}(b_j-b_k)]n_j.
\end{gather}Since it is a spin-singlet, for $a=1,2,3$ the CFT state~\eqref{iMPS} is invariant under global spin rotations $U_a=\exp[i\frac{\pi}{2}S^a_{\mathrm{tot}}]$ by $\pi/2$ around the $x$-, $y$- and $z$-axes. On the subspace of filling fraction $\nu=2/(2m+1)$ and vanishing total spin $S^3_{\mathrm{tot}}=0$ the Halperin infinite MPS for $m\geq 1$ is thus annihilated by the SU(2) invariant operator
\begin{gather}
\notag \frac{1}{6}\sum_{a}U_a^{\dagger}\Big[\sum_{j,\sigma}\big(\Omega'^{\sigma}_j)^{\dagger}\Omega'^{\sigma}_j-m\big(C'^{\sigma}_j)^{\dagger}C'^{\sigma}_j-(m-1)\big(D'^{\sigma}_j)^{\dagger}D'^{\sigma}_j\Big]U_a+\frac{1}{2}(m-1)\sum_{j,\sigma}(Y^{\sigma})^{\dagger}Y^{\sigma}+\frac{1}{2}\tilde{E}_0=\\
\notag (m-1)\sum_{j\neq k}w_{jk}\sum_{\sigma}\dda{j\sigma}\an{k\sigma}-\sum_{j\neq k}w^2_{jk}\Big[-n_j+ \sum_{\sigma}\dda{j\sigma}\an{k\sigma}+\frac{m^2}{2}n_jn_k+m\vec{S}_j\cdot\vec{S}_k\Big]\\
+\sum_{j\neq k}w_{jk}(b_j-b_k)[n_j-\frac{2m+1}{4}n_jn_k-\frac{1}{3}\vec{S}_j\cdot\vec{S}_k]+ \frac{1}{3}m\msum_{j,k,l}w_{jk}w_{jl}\,n_j\vec{S}_k\cdot\vec{S}_l
\end{gather}with a constant $\tilde{E}_0=M(M-1)(M-2)/3+m(m-1)M+M(N-1)(N-2)+(2m+1)M(M-2)/4$. This operator contains only a single three-body term which can be eliminated by adding the SU(2) invariant linear combination
\begin{equation}
\frac{2m}{9}\sum_{a,j}(\Lambda_j^a)^{\dagger}\Lambda_j^a=\sum_{j\neq k}\wb_{jk}w_{jl}[\frac{m}{4}n_jn_k+\frac{m}{3}\vec{S}_j\cdot\vec{S}_k]-\frac{m}{3}\msum_{j,k,l}w_{jk}w_{jl}n_j\vec{S}_k\cdot\vec{S}_l.
\end{equation}In order to obtain the final form of~\eqref{phpbc} one may rewrite $w_{jk}^2=1-(\sin\frac{\varphi_i-\varphi_j}{2})^{-2}$ and simplify the constant terms as
\begin{multline}\label{constterm}
\sum_{j\neq k}\Big[-n_j+\sum_{\sigma}\dda{j\sigma}\an{k\sigma}+\frac{2m^2+m}{4}n_jn_k+m\vec{S}_j\cdot\vec{S}_k]=-(N-1)M\\
+\sum_{\sigma}(Y^{\sigma})^{\dagger}Y^{\sigma}-\ntot
+\frac{2m^2+m}{4}(\ntot-1)\ntot+m\big[(\vec{S}_{\mathrm{tot}})^2-\frac{3}{4}\ntot].
\end{multline}A completely analogous calculation leads to the identity~\eqref{phobc} for the parent Hamiltonian on an open chain $\Gamma=\{u_j=\cos\theta_j|\theta_j\in[0,\pi]\,\forall j=1,\dotsc,N\}$ since the latter is the projection onto the real line of the periodic chain $\{z_j=e^{i\theta_j}|j=1,\dotsc,N\}$ in the upper half plane with complex conjugates $z_{\bar{j}}=e^{-i\theta_j}$ for $j=1,\dotsc,N$ in the lower half plane. The operators in the third column of Tab.~\ref{tab:oppbc} annihilating the infinite MPS on the open chain depend on the lattice sites through $w_{jk}+w_{j\kb}=(z_j+z_k)/(z_j-z_k)+(z_j+\zb_k)/(z_j-\zb_k)$. Due to their close relation with the corresponding expressions on a periodic chain, we have $(w_{ij}+w_{i\bar{j}})(w_{ik}+w_{i\bar{k}})+(w_{ji}+w_{j\bar{i}})(w_{jk}+w_{j\bar{k}})+(w_{ki}+w_{k\bar{i}})(w_{kj}+w_{k \bar{j}})=4$ and $\sum_{i ( \neq j,k)}(w_{ij}+w_{i \bar{j}})(w_{ik}+w_{i \bar{k}})=(4N-6)+2(w^2_{jk}+w^2_{j \bar{k}})+w_{jk}(c_j-c_k)+w_{j \bar{k}}(c_j+c_k)$ for any pairwise different $i,j,k$, where $c_i\equiv w_{\bar{i}i}+\sum_{j(\neq i)}(w_{ji}+w_{\bar{j}i})$. These identities can be used to simplify several terms in the explicit expression for the positive operator $\sum_{j\sigma}(\Omega''^{\sigma}_j)^{\dagger}\Omega''^{\sigma}_j$. Similarly to the periodic case the remaining three-body terms may be absorbed after an explicit SU(2) symmetrisation into $\sum_{a,j,\sigma}U_a^{\dagger}[m(C''^{\sigma}_j)^{\dagger}C''^{\sigma}_j+(m-1)(D''^{\sigma}_j)^{\dagger}D''^{\sigma}_j]U_a/6$ or removed by addition of the operator $(2m/9)\sum_{a,j}(\Lambda''^a_j)^{\dagger}\Lambda''^a_j$. Finally we can rewrite $w^2_{jk}=1-\sin^{-2}\frac{1}{2}(\theta_j-\theta_k)$ to obtain the result~\eqref{phobc}.

\bibliography{halperin}

\end{document}